\documentclass[acmsmall,nonacm,natbib=true]{acmart}

\usepackage{enumitem}
\usepackage{booktabs}
\usepackage{adjustbox}
\usepackage{amsmath,amsfonts}
\usepackage[noend]{algorithmic}
\usepackage[ruled]{algorithm2e}
\usepackage{array}
\usepackage[subrefformat=parens,labelformat=parens]{subfig}
\usepackage{textcomp}
\usepackage{stfloats}
\usepackage{url}
\usepackage{verbatim}
\usepackage{graphicx}
\usepackage{xcolor}
\usepackage{xspace}
\usepackage[acronym]{glossaries}
\usepackage{bbm}
\usepackage{multirow}
\usepackage{tikz}

\newtheorem{definition}{Definition}

\DeclareMathOperator*{\argmax}{arg\,max}
\DeclareMathOperator*{\argmin}{arg\,min}
\DeclareMathOperator*{\E}{\mathbb{E}}
\DeclareMathOperator*{\prob}{\mathbb{P}}
\DeclareMathOperator*{\B}{\mathbb{B}_1^d}
\DeclareMathOperator*{\Var}{Var}

\newcommand{\msmarco}{\textsc{MsMarco v1}\xspace}
\newcommand{\msmarcodos}{\textsc{MsMarco v2}\xspace}
\newcommand{\nq}{\textsc{NQ}\xspace}
\newcommand{\beir}{\textsc{Beir}\xspace}

\newcommand{\seismic}{\textsc{Seismic}\xspace}

\newcommand{\pisa}{\textsc{Pisa}\xspace}
\newcommand{\ioqp}{\textsc{Ioqp}\xspace}
\newcommand{\grassrma}{\textsc{GrassRMA}\xspace}
\newcommand{\pyann}{\textsc{PyAnn}\xspace}
\newcommand{\bruchetal}{\textsc{SparseIvf}\xspace}
\newcommand{\kannolo}{\textsc{kANNolo}\xspace}

\newcommand{\sinnamon}{\textsc{Sinnamon}$_\textsc{Weak}$\xspace}

\newcommand{\splade}{\textsc{Splade}\xspace}
\newcommand{\spladeT}{\textsc{Splade}-v3\xspace}
\newcommand{\esplade}{\textsc{E-Splade}\xspace}

\newcommand{\unicoil}{\textsc{uniCoil-T5}\xspace}

\newcommand{\heapfactor}{\textsf{heap\_factor}\xspace}
\newcommand{\heap}{\textsc{Heap}}

\newcommand{\etal}{\emph{et al.}\xspace}

\newacronym{nns}{\textsc{Nns}}{Nearest Neighbor Search}
\newacronym{anns}{\textsc{Anns}}{Approximate Nearest Neighbor Search}
\newacronym{smips}{Sparse \textsc{Mips}}{Sparse Maximum Inner Product Search}

\newacronym{ts}{\textsc{TS}}{Threshold Sampling}
\newacronym{knn}{\textsc{$\kappa$-nn} graph}{$\kappa$-Nearest Neighbor graph}

\newacronym{sketch}{$\alpha$-MSS}{$\alpha$-Mass Subvector Sketch}
\newacronym{setsketch}{Set $\alpha$-MSS}{Set $\alpha$-Mass Subvector Sketch}
\newacronym{nir}{\textsc{Nir}}{Neural Information Retrieval}
\newacronym[plural=\textsc{Sae}s,firstplural=Sparse Autoencoders (\textsc{Sae}s)]{sae}{\textsc{Sae}}{Sparse Autoencoder}
\newacronym[plural=\textsc{Llm}s,firstplural=Large Language Models (\textsc{Llm}s)]{llm}{\textsc{Llm}}{Large Language Model}

\newcommand{\triacomment}[1]{\hfill \texttt{$\triangleright$ #1}}

\begin{document}

\title[Efficient Sketching and Nearest Neighbor Search Algorithms for Sparse Vector Sets]{Efficient Sketching and Nearest Neighbor Search\\Algorithms for Sparse Vector Sets}

\author{Sebastian Bruch}
\affiliation{%
    \institution{Northeastern University}
    \city{Boston, MA}
    \country{United States of America}
}
\email{s.bruch@northeastern.edu}

\author{Franco Maria Nardini}
\affiliation{%
    \institution{ISTI-CNR}
    \city{Pisa}
    \country{Italy}
}
\email{francomaria.nardini@isti.cnr.it}

\author{Cosimo Rulli}
\affiliation{%
    \institution{ISTI-CNR}
    \city{Pisa}
    \country{Italy}
}
\email{cosimo.rulli@isti.cnr.it}

\author{Rossano Venturini}
\affiliation{%
    \institution{University of Pisa}
    \city{Pisa}
    \country{Italy}
}
\email{rossano.venturini@unipi.it}


\begin{abstract}
Sparse embeddings of data form an attractive class due to their inherent interpretability: Every dimension is tied to a term in some vocabulary, making it easy to visually decipher the latent space. Sparsity, however, poses unique challenges for \gls{anns} which finds, from a collection of vectors, the $k$ vectors closest to a query. To encourage research on this underexplored topic, sparse \gls{anns} featured prominently in a BigANN Challenge at NeurIPS 2023, where approximate algorithms were evaluated on large benchmark datasets by throughput and accuracy. In this work, we introduce a set of novel data structures and algorithmic methods, a combination of which leads to an elegant, effective, and highly efficient solution to sparse \gls{anns}. Our contributions range from a theoretically-grounded sketching algorithm for sparse vectors to reduce their effective dimensionality while preserving inner product-induced ranks; a geometric organization of the inverted index; and the blending of local and global information to improve the efficiency and efficacy of \gls{anns}. Empirically, our final algorithm, dubbed \seismic, reaches sub-millisecond per-query latency with high accuracy on a large-scale benchmark dataset using a single CPU.
\glsresetall
\end{abstract}
\maketitle


\section{Introduction} \label{sec:introduction}

Embeddings have gained increasing popularity since the introduction of \glspl{llm}~\cite{DBLP:series/synthesis/2021LinNY}. These models learn a high-dimensional vector\footnote{We use ``vector'' and ``point'' interchangeably throughout this work.} representation of their input such that the distance between any two input variables is approximately preserved in the target vector space.

As a concrete example, consider \gls{nir} where short text passages are embedded into a vector space, wherein the similarity between two vectors reflects the semantic similarity between the passages they represent. Representing text that way enables a more effective matching of queries to passages by semantic similarity~\cite{INR-071}. That in turn has dramatically changed applications such as web search, recommendation, question answering, and more.

While embedding vectors are typically dense (i.e., each coordinate is almost surely nonzero), there are neural models that produce \emph{sparse} vectors (i.e., in a matrix of embeddings, columns and rows are mostly zero). There are a few reasons why sparse embeddings are enticing, the most important of which is their inherent interpretability.

Sparse embeddings are interpretable because one may ground each dimension in a ``term'' from some fixed ``vocabulary,'' so that by inspecting the nonzero coordinates of an embedding, one can intuitively understand what latent features are extracted from the input and represented in the embedding space. In \gls{nir}, for instance, sparse embedding models~\cite{epic,splade-sigir2021,formal2021splade,formal2022splade,lassance2022efficient-splade} tie each embedding dimension to a word in the source language. A coordinate that is nonzero in an embedding indicates that the corresponding word is semantically relevant to the input.

\subsection{Sparse Maximum Inner Product Search}
Embeddings---dense or sparse---allow us to disentangle data representation from ``search'': the task of finding $k$ dataset items with the largest similarity to an arbitrary query. That is because, if data of any modality (e.g., text, image, video, or audio) are projected into the same vector space, search is simply the problem of \emph{\gls{nns}}, defined formally below:

\begin{definition}[\glsentrylong{nns}]
    \label{definition:nns}
    Given a set $\mathcal{S} \subset \mathbb{R}^d$ with distance function $\delta: \mathbb{R}^d \times \mathbb{R}^d \rightarrow \mathbb{R}$ and query point $q \in \mathbb{R}^d$, \gls{nns} finds the $k$-subset of $\mathcal{S}$ consisting of $k$ closest points to $q$:
    \begin{equation*}
        \argmin_{u \in \mathcal{S}}^{(k)} \delta(q, u).
    \end{equation*}
\end{definition}

For efficiency reasons, however, the problem is often relaxed to \emph{\gls{anns}}, which may erroneously include farther away points in the returned set~\cite{bruch2024foundations}. Its accuracy is quantified as follows, with $\lvert \cdot \rvert$ denoting the size of its argument:

\begin{definition}
    \label{definition:ann-accuracy}
    Suppose $S \subset \mathcal{S}$ is the set of $k$ nearest neighbors of query $q$ in $\mathcal{S}$. The accuracy of an \gls{anns} algorithm that returns a set of $k$ points $S^\prime$, denoted accuracy$@k$, is $\lvert S \cap S^\prime \rvert / k$.
\end{definition}

In this document, we are interested in \emph{inner product} as a measure of \emph{similarity} between vectors. We emphasize that algorithms for \gls{anns} by inner product can be trivially applied to \gls{anns} by cosine similarity, thereby covering the vast majority of use cases. Furthermore, we focus squarely on an instance of the problem over a set of \emph{sparse} vectors, resulting in the following problem:

\begin{definition}[\gls{smips}]
    \label{definition:mips}
    Consider a set $\mathcal{S} \subset \mathbb{R}^d$ of sparse vectors and query point $q$. \gls{smips} is an instance of \gls{nns} that finds the subset of $k$ points in $\mathcal{S}$ most similar to $q$ by inner product, denoted by $\langle \cdot, \cdot \rangle$:
    \begin{equation*}
        \argmax_{u \in \mathcal{S}}^{(k)} \langle q, u \rangle.
    \end{equation*}
\end{definition}

\subsection{Challenges in Sparse \textsc{Mips}}

While Dense and \gls{smips} are conceptually the same, sparsity poses unique challenges. First, any computation involving sparse vectors is incompatible with widely-adopted modern platforms (e.g., Graphics Processing Units and Tensor Processing Units). That is because such hardware platforms are designed to process data that is represented as contiguous arrays of floating point numbers---the standard way to store and represent dense vectors. While it is possible to represent sparse vectors in that format, that would be impractical, wasteful, and ultimately ineffective in high dimensions---often in the order of tens to hundreds of thousands of dimensions.

Second is a dearth of efficient representations. Contrast that to dense vectors for which a plethora of algorithms exist to reduce the amount of space taken up by a vector while (a) approximately preserving distances and (b) enabling fast distance computation~\cite{pq,opq,pqWithGPU,locallyOptimizedPQ,multiscaleQuantization,scann,woodruff2014sketching}.

The two challenges above---computational and storage inefficiencies---make \gls{smips} far more resource-intensive than the same operation over dense vector sets. That is further exacerbated as datasets grow larger in the number of vectors or the number of nonzero coordinates per vector.

To counter some of these issues, a number of efforts have explored specialized algorithms for \gls{smips}~\cite{bruch2023sinnamon,bruch2023bridging,formal2023tois-splade,10.1145/3576922,mallia2022guided-traversal}. In fact, to encourage the community to design practical algorithms, the 2023 BigANN Challenge~\cite{simhadri2024resultsbigannneurips23} at the Neural Information Processing Systems (NeurIPS) conference dedicated a track to this problem.

\subsection{\seismic: Our Proposed Algorithm}

Inspired by BigANN, we presented in~\cite{bruch2024seismic} a novel \gls{smips} algorithm that we called \seismic (\textbf{S}pilled Clust\textbf{e}ring of \textbf{I}nverted Lists with \textbf{S}ummaries for \textbf{M}aximum \textbf{I}nner Produ\textbf{c}t Search). The algorithm, which we describe in detail in Section~\ref{sec:algorithm}, admits effective and efficient \gls{smips}.

\seismic is realized by the amalgamation of a set of novel data structures and algorithmic contributions, each of which may be of independent interest. One such contribution is an elegant sketching algorithm to reduce the effective dimensionality of sparse vectors into a lower effective dimensionality such that inner products are approximately preserved with high probability. This sketching algorithm, which we call \gls{setsketch}, is critical to the success of \seismic. We describe this in more detail and present an extensive analysis in Section~\ref{section:sketch}.

Another novel contribution is a data structure. Rather, \seismic \emph{enhances} an existing data structure called the \emph{inverted index} (i.e., a mapping from dimensions to an \emph{inverted list} containing vectors with a nonzero coordinate in that dimension). The enhancement is through the organization of inverted lists into geometrically-cohesive blocks---an approach previously explored for other tasks~\cite{Cinar2023ClusterSkipping,Altingovde2008ClusterSkipping}. Each block is then summarized by a representative point with the idea that the summaries help identify the subset of blocks that can be dismissed without evaluation during search. This logic turns out to be a critical factor in enabling highly efficient search.

Our evaluation of \seismic against state-of-the-art baselines in~\cite{bruch2024seismic}, which included the top (open-source) submissions to the BigANN Challenge, empirically demonstrated that average per-query latency was in the \emph{microsecond} territory on various sparse embeddings of \msmarco~\cite{nguyen2016msmarco}. Impressively, \seismic outperformed the winning solutions of the BigANN Challenge by a factor of at least $3.4\times$ at $95\%$ accuracy@$10$ on \splade---a sparse embedding model---and $12\times$ on Efficient \splade---another sparse embedding model---with the margin widening substantially as accuracy increases. Other baselines were consistently \emph{one to two orders of magnitude} slower than \seismic.

\medskip

In a follow-up work~\cite{Bruch2024SeismicWave}, we further improved \seismic's query processing logic. We did so by incorporating another data structure that dovetails with the inverted index from before. The structure is the \gls{knn}, a directed graph where each data point is connected to its $\kappa$ nearest neighbors.

When the inverted index returns an approximate set of nearest neighbors, we use the \gls{knn} to expand that set and form a larger pool of possible nearest neighbors. We do so by taking each point in the returned set, collecting the points that are one hop away from it in the \gls{knn}, and inserting the new points into the larger set. The expanded set is then re-evaluated and the closest $k$ points are returned as the (refined) approximate nearest neighbors.

The reason this procedure improves the accuracy of \gls{smips} is that it blends the ``global'' structure of the embedding space (captured by the inverted index) with the ``local'' structure (captured by the graph). This complementarity is not new in and of itself; it has guided prior research~\cite{10.1145/3511808.3557231,10.1145/3539618.3591715}. Our experiments reported in~\cite{Bruch2024SeismicWave}, confirm the intuition above: given a fixed memory budget, providing the \gls{knn} to \seismic improves latency by up to a factor of $2\times$ for the same level of accuracy. This was especially advantageous if a near perfect accuracy is desired.

Finally, in another empirical study~\cite{ecir-seismic}, we strengthened our claims by applying \seismic to a large-scale dataset of about $140$ million sparse embedding vectors. As we reported in that work, \seismic scales well to such massive datasets: it indexes that collection in a fraction of the time required by other algorithms; and, it produces $10$ approximate nearest neighbors per query point in a mere $3$ milliseconds with $98\%$ accuracy.

\subsection{Contributions}

This work consolidates and extends our contributions from~\cite{bruch2024seismic,Bruch2024SeismicWave,ecir-seismic} and presents a comprehensive evaluation of the final algorithm on benchmark datasets. Concretely, we make the following contributions in this work:

\begin{itemize}[leftmargin=*]
    \item We present the evolution of \gls{setsketch}, our sketching algorithm for sparse vectors, along with a comprehensive theoretical analysis. This analysis, which explains the choices we made in the design and implementation of the algorithm, was not included in our previous publications.
    
    \item We present a unified algorithm from the pieces scattered in separate publications, and tie it back to our theoretical analysis.

    \item We extend our empirical evaluation of \seismic to include new embedding models and datasets; compare against other baseline \gls{smips} algorithms; and, examine the sensitivity of \seismic to its hyperparameters and study the algorithm's generalizability.
\end{itemize}

\subsection{Outline} The remainder of this paper is organized as follows. Section~\ref{sec:related} reviews related work to position our work in the literature. In Section~\ref{section:sketch}, we give a detailed account of our sketching algorithm with theoretical analysis. The main algorithm, \seismic, is then described in Section~\ref{sec:algorithm}. We evaluate \seismic in Section~\ref{section:experiments} and conclude our work in Section~\ref{sec:conclusions}.

\section{Related Work} \label{sec:related}

This section reviews notable related research. We first summarize the thread of work on sparse embedding vectors by highlighting their use cases. We then venture into the field of text retrieval and review the algorithms presented in that literature. The reason for the latter topic is that we borrow the inverted index data structure from that literature---although as we explained earlier, we enhance it in novel ways and adapt it to \gls{smips}.

\subsection{Sparse Embedding Vectors}

\begin{figure}[t]
    \centering
    \subfloat[\gls{nir}]{
        \includegraphics[width=0.46\linewidth]{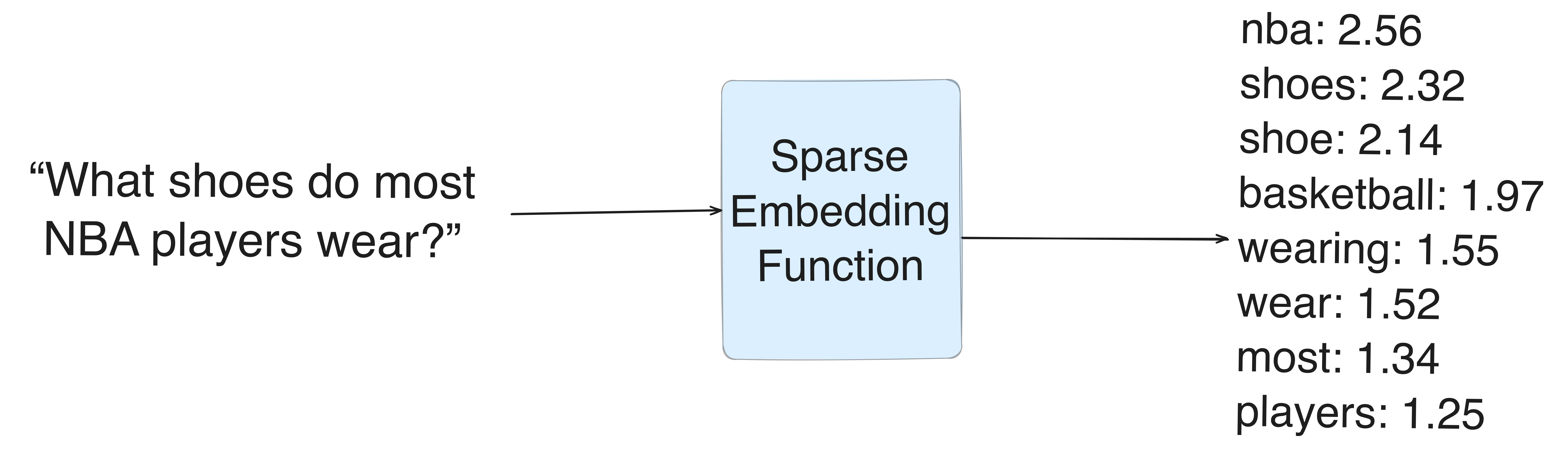}
        \label{fig:sparse-vector-applications:text-retrieval}
    }
    \subfloat[Interpretability]{
        \includegraphics[width=0.46\linewidth]{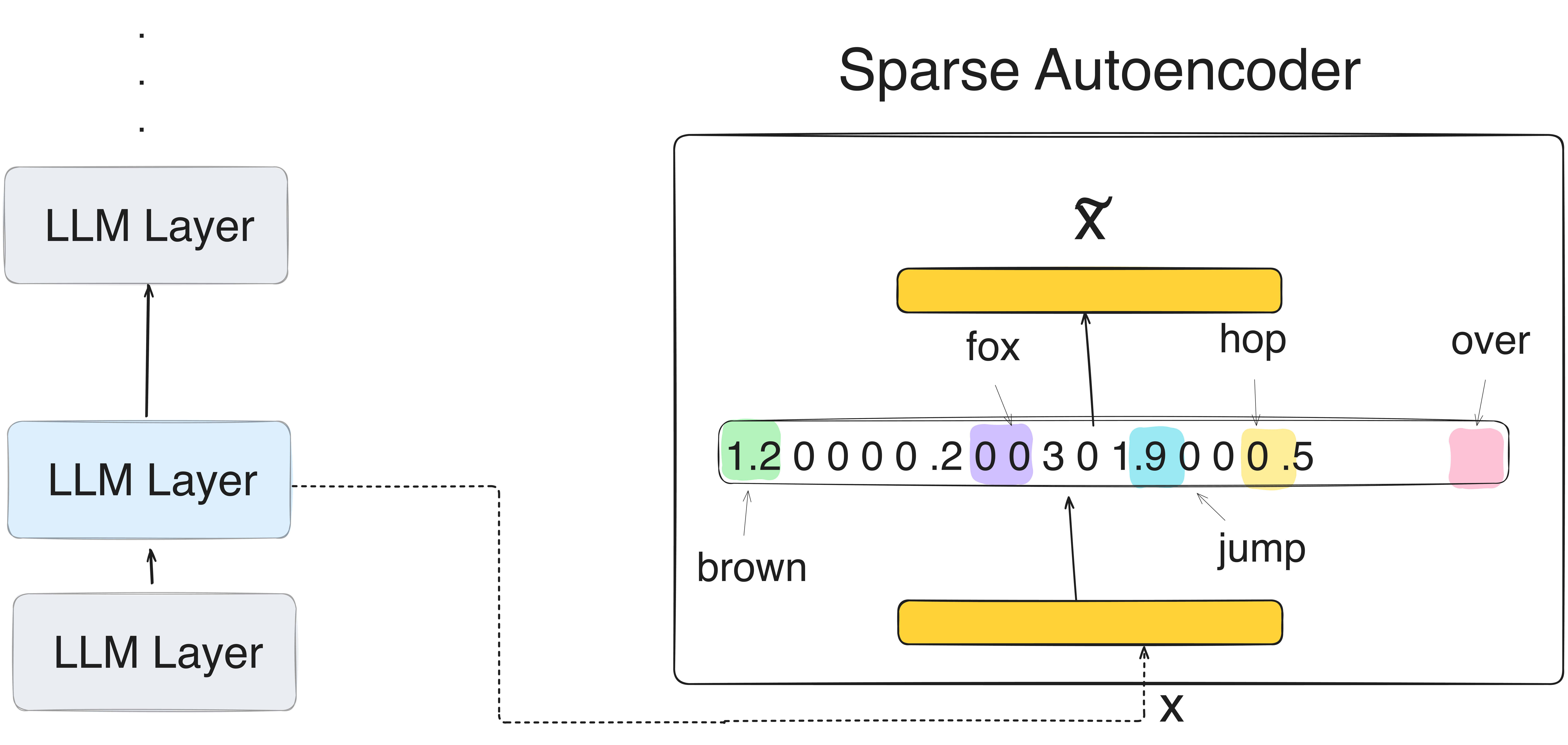}
        \label{fig:sparse-vector-applications:sae}
    }
    \caption{Example usage of sparse embeddings. In (a), a text input is embedded into a sparse vector (shown as a set of nonzero coordinate-value pairs), where terms that are semantically similar to the input have nonzero values. Such embeddings enable accurate and interpretable query-document matching in text retrieval. In (b), a Sparse Autoencoder probes one layer of an \gls{llm}, translating its dense output into a sparse vector and back (such that $x \approx \tilde{x}$). The sparse vector reveals the features that the layer has learned, shedding light on what that layer processes.}
    \label{fig:sparse-vector-applications}
\end{figure}

Sparse embeddings were investigated in the context of text retrieval long before \glspl{llm}~\cite{zamani2018sparse}. But the rise of \glspl{llm} supercharged this research and led to a flurry of activity on the topic~\cite{dai2019contextaware,10.1145/3366423.3380258,10.1145/3397271.3401204,epic,zhao-etal-2021-sparta,sparterm,formal2021splade,formal2023tois-splade,unicoil}. See Figure~\subref*{fig:sparse-vector-applications:text-retrieval} for an example sparse embedding of text used in text retrieval.

What the cited references claim and show empirically is that \gls{smips} over their proposed sparse embeddings (e.g., \splade~\cite{formal2021splade}) can reach state-of-the-art performance for the task of text retrieval on a suite of benchmark datasets like \msmarco{}~\cite{nguyen2016msmarco} and \beir{}~\cite{thakur2021beir}. All that while also lending interpretability to the embeddings of text documents and queries, as argued earlier.

Unsurprisingly, due to their highly interpretable nature, sparse embeddings also play a pivotal role in the literature on interpreting \glspl{llm}. That literature is peppered with algorithms that ``probe'' an \gls{llm} by way of learning a sparse representation of its layers.

The tool that does the probing is known as a \gls{sae}~\cite{oneill2024sparseautoencodersenablescalable,cunningham2023sparseautoencodershighlyinterpretable}. As illustrated in Figure~\subref*{fig:sparse-vector-applications:sae}, \glspl{sae} learn to project a dense vector (obtained from an \gls{llm}'s internal layer) to a sparse vector, then back to the same dense vector space. The learning is unsupervised: \glspl{sae} learn to reconstruct their input. What is interesting, however, is the sparse embedding that is learned in the middle: It is argued that this projection ``reverses'' superposition~\cite{elhage2022toymodelssuperposition} and polysemanticity~\cite{cunningham2023sparseautoencodershighlyinterpretable}, allowing us to ``find concise, human-understandable explanations for what neural networks are doing internally''~\cite{cunningham2023sparseautoencodershighlyinterpretable,oneill2024sparseautoencodersenablescalable,braun2024identifyingfunctionallyimportantfeatures,marks2025sparse,he2024llamascopeextractingmillions,lieberum-etal-2024-gemma}. In this context, \gls{smips} can be used to gather insight not from individual vectors, but a large collection of them.

\subsection{Text Retrieval Algorithms}

The information retrieval literature offers a wide array of algorithms tailored to retrieval on text collections~\cite{tonellotto2018survey}: given a text query, the task is to find the top $k$ text documents that are ``relevant'' to the query. Theses algorithms are often \emph{exact} and scale easily to massive datasets. MaxScore~\cite{maxscore} and WAND~\cite{broder2003wand}, and their intellectual descendants~\cite{ding2011bmwand,topk_bmindexes,mallia2019faster-blockmaxwand,mallia2017blockmaxwand_variableBlocks}, are examples that perform search over a ``bag-of-words'' representation of text, such as BM25~\cite{bm25original} or TF-IDF~\cite{salton1988term}.

These algorithms operate on an inverted index, augmented with additional data to speed up query processing. One that features prominently is the maximum attainable partial inner product---an upper-bound. This enables the possibility of navigating the inverted lists, one document at a time, and deciding quickly if a document may belong to the result set. Effectively, such algorithms (safely) \emph{prune} the parts of the index that cannot be in the top-$k$ set. That is why they are often referred to as \emph{dynamic pruning} techniques.

Although efficient in practice, dynamic pruning methods are designed specifically for text collections. Importantly, they ground their performance on several pivotal assumptions: nonnegativity, higher sparsity rate for queries, and a Zipfian shape of the length of inverted lists. These assumptions are valid for TF-IDF or BM25, which is the reason why dynamic pruning works well and the worst-case time complexity of search is seldom encountered in practice.

These assumptions do not necessarily hold for sparse embeddings, however. Embedding vectors may be real-valued, with a sparsity rate that is closer to uniform across dimensions~\cite{bruch2023sinnamon,mackenzie2021wacky}. Mackenzie \etal~\cite{10.1145/3576922} find that sparse embeddings reduce the odds of pruning or early-termination in the document-at-a-time (DaaT) and Score-at-a-Time (SaaT) paradigms. That is not to say there have not been attempts to modify inverted index-based algorithms to make them compatible with sparse embeddings~\cite{mallia2022guided-traversal,mallia2024faster,mackenzie-etal-2022-accelerating}. Unlike these \emph{exact} methods, our approach is \emph{approximate} by design, offering far greater flexibility to sacrifice accuracy for latency per operational requirements.

The most similar work to ours is~\cite{bruch2023bridging}. The authors investigate if \gls{anns} algorithms for \emph{dense} vectors port over to \emph{sparse} vectors. They focus on \emph{inverted file} (IVF) where vectors are partitioned into clusters during indexing, with only a fraction of clusters scanned during search. They show that IVF serves as an efficient solution for sparse \gls{anns}. Interestingly, the authors cast IVF as dynamic pruning and turn that insight into a novel organization of the inverted index for sparse \gls{anns} for general sparse vectors. Our index structure can be viewed as an extension of theirs.

Finally, we briefly describe another \gls{anns} algorithm over dense vectors: HNSW~\cite{hnsw2020}, an algorithm that constructs a graph where each data point is a node and two nodes are connected if they are deemed ``similar.'' Similarity is based on Euclidean distance, but~\cite{ip-nsw18} shows inner product too can be used to form the graph (known as IP-HNSW). As we learn in the presentation of our empirical analysis, algorithms that adapt IP-HNSW~\cite{ip-nsw18} to sparse vectors work remarkably well.

\glsreset{sketch}

\section{Sparse Vector Sketching}
\label{section:sketch}

As we noted earlier, one of the challenges with sparse vectors that have a large number of nonzero coordinates is that exact \gls{nns} algorithms struggle to perform the task when the latency budget is tight. To make search efficient, we explore a method to reformulate the problem as \gls{smips}.

The cornerstone of our reformulation is a sketching algorithm that reduces the number of nonzero coordinates such that the relative order of two points induced by their inner product with a query is approximately preserved. This section presents our sketching algorithm, which we call \gls{setsketch}, along with a thorough analysis.

\subsection{Notation}
Before we describe the algorithm, let us introduce our notation. We use lower-case letters (e.g., $u$) to denote a vector and write $u_i$ for its $i$-th coordinate. A sparse vector can be identified as a set of nonzero entries $\{ (i, u_i) \;|\; u_i \neq 0 \}$.

Let $\B = \{ u \in [0, 1]^d \;|\; \lVert u \rVert_1 \leq 1 \}$ be the nonnegative region of the $\ell_1$ unit ball. Suppose we have a set $\mathcal{S} \subset \B$ of nonnegative \emph{sparse} vectors. We assume $\mathcal{S} \sim \mathcal{D}$ for some unknown data distribution $\mathcal{D}$. We use superscript to enumerate: $u^{(j)}$ is the $j$-th vector in $\mathcal{S}$.

Let us define a few concepts that we frequently refer to. The $\ell_p$ norm of a vector denoted by $\lVert \cdot \rVert_p$ is defined as $\lVert x\rVert_p = (\sum_i \lvert x_i\rvert^p)^{1/p}$. We call the $\ell_p$ norm of a vector its $\ell_p$ \emph{mass}. Additionally, we use the notion of a \emph{subvector} throughout this work: The subvector $\tilde{u}$ of sparse vector $u$ is a vector with the property that $\tilde{u}_i \neq 0 \implies u_i \neq 0 \;\land\; \tilde{u}_i = u_i$. Finally, a vector $u$ \emph{restricted to a set of dimensions} $\mathcal{I}$ is a vector $u_\mathcal{I}$ such that $(u_\mathcal{I})_i = u_i$ if $i \in \mathcal{I}$ and $0$ otherwise.

\subsection{$\ell_1$ Threshold Sampling}
A recent study~\cite{daliri2023sampling} presented a sketching algorithm for sparse vectors that rests on the following principle: Coordinates that contribute more heavily to the $\ell_2$ norm squared weigh more heavily on the inner product between vectors. Based on that intuition, they report that if we were to drop the nonzero coordinates of a sparse vector with a probability proportional to their contribution to the $\ell_2$ norm squared, we can reduce the size of vectors while approximately maintaining their inner products. We refer to their sketching algorithm as $\ell_2$-\gls{ts}.

\subsubsection{The $\ell_1$-\gls{ts} algorithm}
We build on that same intuition, but consider instead the $\ell_1$ norm. We make this choice because, on the datasets we use in our experiments, $\ell_1$-\gls{ts} yields inner product estimates that are far more accurate than the $\ell_2$ variant. Our algorithm detailed in Algorithm~\ref{algorithm:l1-ts} is similar to the $\ell_2$-\gls{ts} procedure and enjoys similar properties. Given the $\ell_1$-\gls{ts} sketches of two vectors $u$ and $v$, their inner product is estimated as follows:
\begin{equation}
    \label{equation:l1-ts-ip}
    \sum_{i \in K_u \cap K_v} \frac{u_i v_i}{\min(1, u_i \tau_u, v_i \tau_v)},
\end{equation}
where $K_u$, $K_v$, $\tau_u$, and $\tau_v$ are produced by Algorithm~\ref{algorithm:l1-ts}.

\begin{algorithm}[t]
    \caption{$\ell_1$ Threshold Sampling}
    \label{algorithm:l1-ts}
    \textbf{Input}: $u \in \B$; target sketch size $d_\circ$; uniformly random hash function $h: \{1,\ldots,d\} \rightarrow [0, 1]$.
    
    \KwResult{Sketch $ts(u) = \{K_{u}, V_{u}, \tau_{u}\}$, where $K_{u}$ is a subset of dimensions from $\{1, \ldots, d\}$; $V_{u}$ contains $u_i$ for all $i\in K_{u}$; and, $\tau_{u} \in \mathbb{R}_+$.}
    \begin{algorithmic}[1]
        \STATE $K_{u} \leftarrow \emptyset; V_{u} \leftarrow \emptyset$
        \FOR{$i$ such that $u_i \neq 0$}
            \STATE $\tau_i \leftarrow  d_\circ \cdot \frac{u_i}{\lVert u \rVert_1}$.
            \IF{$h(i) \leq \tau_i$}
              \STATE $K_u \leftarrow K_u \cup \{ i \}; V_u \leftarrow V_u \cup \{ u_i \}$
            \ENDIF
        \ENDFOR
        \RETURN $ts(u) = \{K_{u}, V_u, \tau_u\}$ where $\tau_u = d_\circ/\lVert u \rVert_1$.
    \end{algorithmic}
\end{algorithm}

\subsubsection{Analysis of $\ell_1$-\gls{ts}}

\begin{theorem}\label{theorem:l1-ts}
For vectors $u, v \in \B$ and sketch size $d_\circ$, let $ts(u)$ and $ts(v)$ denote sketches produced by Algorithm~\ref{algorithm:l1-ts}. Let $W$ be the inner product estimated by Equation~\eqref{equation:l1-ts-ip}. We have that $\E\left[W\right] = \langle u, v \rangle$ and
\begin{equation*}
\Var[W] \leq \frac{1}{d_\circ} \langle u, v \rangle \big( \lVert u \rVert_1 + \lVert v \rVert_1 \big).
\end{equation*}
\end{theorem}
\begin{proof}
    Let $\mathcal{I}$ denote the set of all indices $i$ for which $u_i \neq 0$ and $v_i \neq 0$. For any $i\in \mathcal{I}$, let $\mathbbm{1}_i$ denote the indicator random variable for the event that $i \in K_u \cap K_v$. Note that, for $i\neq j$, $\mathbbm{1}_i$ is independent from $\mathbbm{1}_j$, because the hash values $h(i)$ and $h(j)$ are drawn independently. Now, $\prob[\mathbbm{1}_i = 1]$ over all hash functions is equal to:
    \begin{equation*}
        \label{equation:prob_claim}
        p_i = \min\left(1, \frac{d_\circ u_i}{\lVert u \rVert_1}, \frac{d_\circ v_i}{\lVert v \rVert_1}\right) = \min(1, \tau_u u_i, \tau_v v_i).
    \end{equation*}

    Considering the independence of $\mathbbm{1}_i$'s, we can write $W$ as $\sum_{i\in \mathcal{I}} \mathbbm{1}_i \cdot u_i v_i/p_i$. By linearity of expectation, we have that:
    \begin{equation*}
        \E[W] = \sum_{i\in \mathcal{I}} p_i \frac{u_i v_i}{p_i} = \sum_{i\in \mathcal{I}} u_i v_i = \langle u, v\rangle.
    \end{equation*}

    As for the variance, we again take advantage of independence and write:
    \begin{align*}
    \Var[W] &\leq 
    \frac{1}{d_\circ} \sum_{i\in \mathcal{I},\; p_i \neq 1} \frac{u_i^2 v_i^2 }{ \min( u_i / \lVert u \rVert_1, v_i / \lVert v \rVert_1)}
    = \frac{1}{d_\circ} \sum u_i^2 v_i^2 \max( \lVert u \rVert_1 / u_i, \lVert b \rVert_1 / v_i) \\
    &= \frac{1}{d_\circ}\sum \max( u_i v_i^2 \lVert u \rVert_1, u_i^2 v_i \lVert v \rVert)
    \leq \frac{1}{d_\circ} \sum u_i v_i^2 \lVert u \rVert_1 + u_i^2 v_i \lVert v \rVert_1 \\
    &= \frac{1}{d_\circ} \big( \lVert u \rVert_1 \sum u_i v_i^2 + \lVert v \rVert_1 \sum u_i^2 v_i \big)
    \leq \frac{1}{d_\circ} \big( \lVert u \rVert_1 \langle u, v \rangle + \lVert v \rVert_1 \langle u, v \rangle \big)  \\
    &= \frac{1}{d_\circ} \langle u, v \rangle \big( \lVert u \rVert_1 + \lVert v \rVert_1 \big).
    \end{align*}
\end{proof}

By applying the one-sided Chebyshev's inequality, we can derive the following corollary.
\begin{corollary}
For all values of $\epsilon, \delta \in (0, 1)$ and sparse vectors $u, v \in \B$, $\ell_1$-\gls{ts} with sketch size $d_\circ$ returns inner product $W$ such that, with probability at least $1 - \delta$:
\begin{equation*}
    W - \langle u, v \rangle \leq \sqrt{\frac{1 - \delta}{\delta d_\circ} \langle u, v \rangle \big( \lVert u \rVert_1 + \lVert v \rVert_1 \big)}.
\end{equation*}
\end{corollary}

\subsection{\glsentrylong{sketch}}
The problem with Algorithm~\ref{algorithm:l1-ts} is that the inner product of two sketches is a \emph{weighted} product of surviving entries. That weight is a function of \emph{both} vectors. That implies, when one of the vectors is a dynamic query point $q$, the weights have to be calculated for every data point $u$ during search, making \gls{smips} complex and time consuming.

In this section, we use the principle underlying $\ell_1$-\gls{ts}, but make its stochastic nature deterministic and simplify the inner product estimate computation. Doing so has an important caveat: the inner product of two sketches is an \emph{under-estimate} of the true inner product. But as we will see later, this modification has its own advantages.

\subsubsection{The sketching algorithm}

\begin{definition}[$\alpha$-mass subvector sketch]
\label{definition:alpha-mss}
Consider a vector $u \in \B$ and a constant $\alpha \in (0, 1]$. Take a permutation $\pi$ such that $\lvert u_{\pi_i} \rvert \geq \lvert u_{\pi_{i+1}} \rvert$ and denote by $j$ the smallest integer such that:
\begin{equation*}
    \sum_{i=1}^j \lvert u_{\pi_i} \rvert \geq \alpha \lVert u \rVert_1.
\end{equation*}
We call $\tilde{u}$ made up of $\{ (\pi_i, u_{\pi_i}) \}_{i=1}^j$, the \gls{sketch} of $u$. Furthermore, the inner product estimate of two sketches $\tilde{u}$ and $\tilde{v}$ is defined simply as $\langle \tilde{u}, \tilde{v} \rangle = \sum \tilde{u}_i \tilde{v}_i$.
\end{definition}

It is easy to see that the sketch of Definition~\ref{definition:alpha-mss} leads to \emph{biased} inner product estimates. However, given an \gls{sketch} of $u$ and a $\beta$-MSS of $v$, we can present the following bound on the error of the estimated inner product.

\begin{theorem}
    \label{theorem:alpha-mss-bound}
    Consider an \gls{sketch} $\tilde{u}$ of $u$ and $\beta$-MSS $\tilde{v}$ of $v$, where $u, v \in \B$. Defining $\mathcal{I} = \{ i \; \mathtt{s.t.}\; u_i \neq 0 \;\land\; v_i \neq 0 \}$, the following bound holds:
    \begin{equation*}
        \langle u,\; v \rangle - \langle \tilde{u},\; \tilde{v} \rangle
        \leq (1 - \alpha)\lVert u \rVert_1 \cdot (1 - \beta)\lVert v \rVert_1 +
        \lVert u_\mathcal{I} \rVert_1 \cdot (1 - \beta) \lVert v \rVert_1 +
        \lVert v_\mathcal{I} \rVert_1 \cdot (1 - \alpha) \lVert u \rVert_1,
    \end{equation*}
    where $u_\mathcal{I}$ is the vector $u$ restricted to $\mathcal{I}$.
\end{theorem}
\begin{proof}
    Let $\mathbbm{1}_p$ denote the indicator function which is $1$ if the predicate $p$ holds.
    \begin{align*}
        \langle u,\; v \rangle - \langle \tilde{u},\; \tilde{v} \rangle
        &= \sum_{i \in \mathcal{I}} u_i v_i - \tilde{u}_i \tilde{v_i}
        = \sum_{i \in \mathcal{I}} \mathbbm{1}_{\tilde{u_i} = 0 \;\lor\; \tilde{v}_i = 0} \cdot u_i v_i \\
        &= \sum_{i \in \mathcal{I}} \mathbbm{1}_{\tilde{u_i} = 0 \;\land\; \tilde{v}_i = 0} \cdot u_i v_i + \mathbbm{1}_{\tilde{u_i} = 0 \;\land\; \tilde{v}_i \neq 0} \cdot u_i v_i + \mathbbm{1}_{\tilde{u_i} \neq 0 \;\land\; \tilde{v}_i = 0} \cdot u_i v_i.
    \end{align*}
    Note that $\lVert \tilde{u} \rVert_1 \geq \alpha \lVert u \rVert_1$, $\lVert \tilde{v} \rVert_1 \geq \beta \lVert v \rVert_1$, and $\sum u_i v_i \leq \lVert u \rVert_1 \lVert v \rVert_1$. As such the first term is bounded above by $\sum_{i \in \mathcal{I}} \mathbbm{1}_{\tilde{u}_i = 0} u_i \cdot \sum_{i \in \mathcal{I}} \mathbbm{1}_{\tilde{v}_i = 0} v_i$, which in turn is bounded by $(1 - \alpha)\lVert u \rVert_1 \cdot (1 - \beta)\lVert v \rVert_1$. The last two terms follow a similar reasoning. That yields the desired bound.
\end{proof}

\subsubsection{Empirical evidence}
Let us demonstrate the result above on the \msmarco dataset~\cite{nguyen2016msmarco}---a benchmark text collection to be described in more detail later in this work---embedded into a sparse vector space with \splade{}~\cite{formal2022splade}.\footnote{The \texttt{cocondenser-ensembledistill} checkpoint was obtained from \url{https://huggingface.co/naver/splade-cocondenser-ensembledistil}.} We take every query and data vector, sort its entries by coordinate in descending order, and measure the fraction of the $\ell_1$ mass preserved by considering a given number of top coordinates.

\begin{figure}[t]
\centering
\subfloat[]{
    \label{figure:alpha-mss:evidence:mass}
    \includegraphics[width=0.49\columnwidth]{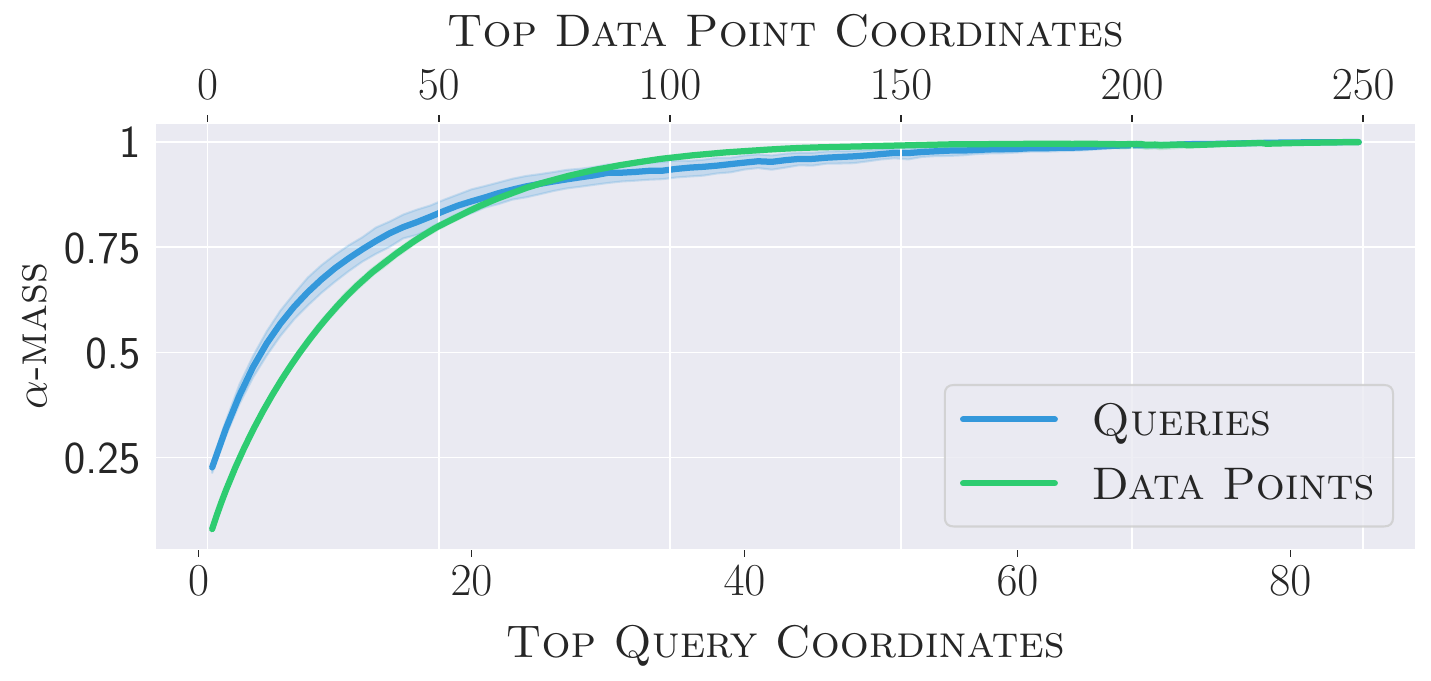}
}
\subfloat[]{
    \label{figure:alpha-mss:evidence:ip}
    \includegraphics[width=0.49\columnwidth]{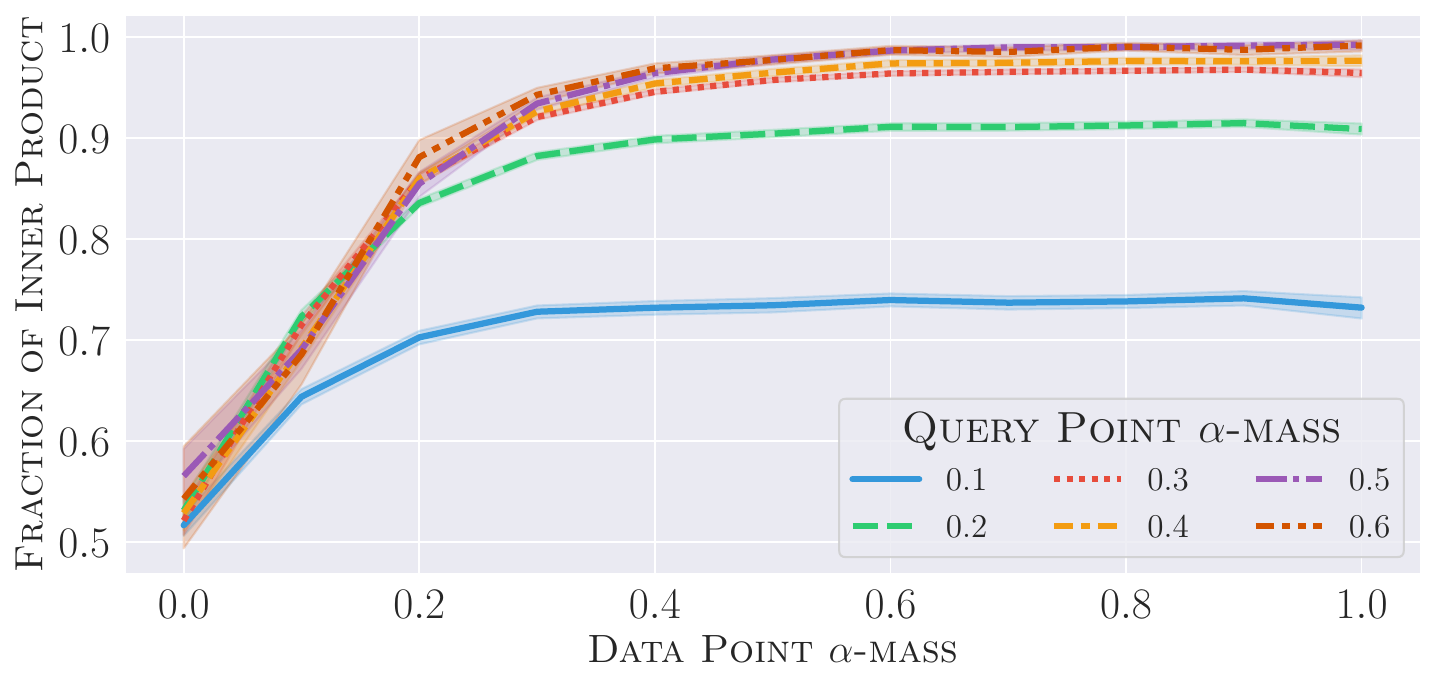}
}
\caption{Empirical demonstration of \gls{sketch} on the \splade embeddings of the \msmarco dataset: (a) Fraction of $\ell_1$ mass preserved by keeping only the largest nonzero entries. This shows the number of coordinates that must be preserved (horizontal axis) in order to construct an $\alpha$-mass subvector (vertical axis). (b) Fraction of true inner product (with $95\%$ confidence intervals) preserved by \gls{sketch}.
\label{figure:alpha-mss:evidence}}
\end{figure}

For query points, the top $10$ entries yield $0.75$-mass subvectors. For data points, the top $50$ (about $30$\% of nonzero entries), give $0.75$-mass subvectors. We illustrate our measurements in Figure~\subref*{figure:alpha-mss:evidence:mass}. Next, consider the inner product estimate using \gls{sketch} between query and data points. This is shown in Figure~\subref*{figure:alpha-mss:evidence:ip}. We observe that, a $0.2$-MSS sketch of queries and documents preserves an average of $85\%$ of the true inner product.

\subsection{\glsentrylong{setsketch}}
\label{section:sketch:global-mss}

The sketching algorithms presented thus far apply to a vector independently of other vectors. As such, given a collection of vectors $\mathcal{S}$ and a constant $\alpha$, the sketch of every vector $u \in \mathcal{S}$ has a mass that is at least $\alpha \lVert u \rVert_1$. While that may be desirable for datasets where every vector has an equal likelihood of maximizing the inner product with an arbitrary query $q$, it becomes wasteful for datasets where that likelihood across vectors is non-uniform.

Let us elaborate by considering an extreme case. Suppose $u, v \in \B$ are such that $u = t v$ for some $t \in [0, 1)$. It is clear that, $\langle q, u \rangle < \langle q, v \rangle$ for any $q \in \B$. If $u$ and $v$ belong to the set $\mathcal{S}$, and noting that our ultimate goal is to find the solution to \gls{smips} over $\mathcal{S}$ for $q$, then it is not necessary for the sketches of $u$ and $v$ to preserve the same fraction $\alpha$ of their $\ell_1$ mass. Instead, it makes more sense for $\alpha$ to \emph{adapt} to the likelihood of $u$ and $v$ being the solution to \gls{smips}: if $\tilde{v}$ is the $\alpha_v$-MSS of $v$ and $\tilde{u}$ is the $\alpha_u$-MSS of $u$, $\alpha_u < \alpha_v$ is a reasonable choice.

\begin{figure}[t]
    \centering
    \includegraphics[width=0.6\linewidth]{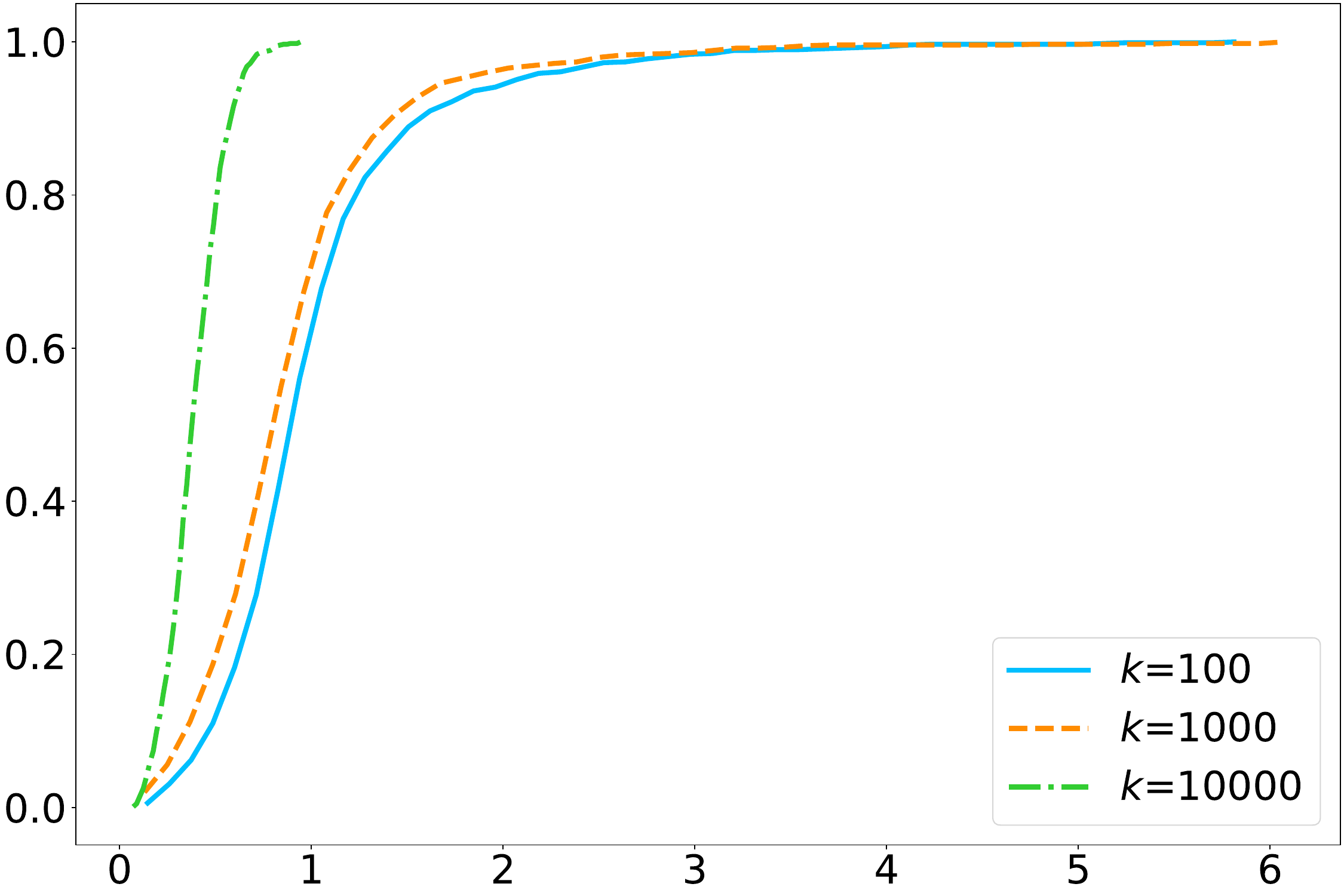}
    \caption{We sample $1{,}000$ queries from \msmarco, find their nearest neighbor (say, $u$) and the $k$-th nearest neighbor (say, $v$) for some large $k > 1$. We then compute the ratio of their $\ell_1$ norms, $\lVert v_\mathcal{I} \rVert_1 / \lVert u_\mathcal{I} \rVert_1$, where $\mathcal{I}$ is the set of active dimensions in the query. The figure shows the cumulative distribution (vertical axis) of this ratio (horizontal axis). As $k$ increases, so that we include farther points, the ratio becomes smaller, showing that points that are closer to the query have a larger $\ell_1$ norm along the active dimensions of the query.}
    \label{figure:bound_ratios}
\end{figure}

While the extreme example above rarely occurs, the property that some vectors are more likely to be the solution of \gls{smips} than others appears to hold in many practical datasets and query distributions. For instance, consider the benchmark \msmarco dataset and its queries: nearest neighbors of each query generally have a larger $\ell_1$ norm along a query's active dimensions (i.e., dimensions corresponding to nonzero coordinates) than father neighbors. To demonstrate this, we sample $1{,}000$ queries from the dataset, find their nearest neighbor as well as the $k$-th nearest neighbor for some large $k > 1$. We then compute and plot in Figure~\ref{figure:bound_ratios} the ratio of the $\ell_1$ norm of the two neighbors. The plot clearly shows this phenomenon.

In this section, we wish to sketch the entire data matrix (i.e., the set $\mathcal{S}$) at once such that each vector is sketched in the context of other vectors in an adaptive manner. In particular, on the basis of the empirical observation above, we wish the amount of mass preserved for each vector to depend on the magnitude of its $\ell_1$ mass relative to other vectors'. The remainder of this section describes our sketching algorithm, which we call \glsentrylong{setsketch}.

\subsubsection{\glsentrylong{setsketch}}
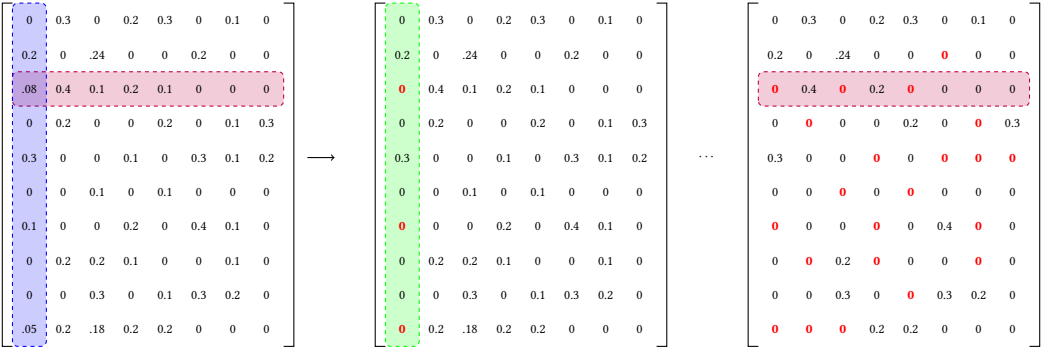
\begin{figure}[t]
\centering
\resizebox{\textwidth}{!}{
\begin{tikzpicture}[every node/.style={minimum size=4mm, anchor=center}, scale=0.9]

\def\rows{10}
\def\cols{8}

\def\matrixA{
  {0, 0.3, 0, 0.2, 0.3, 0, 0.1, 0},
  {0.2, 0, .24, 0, 0, 0.2, 0, 0},
  {.08, 0.4, 0.1, 0.2, 0.1, 0, 0, 0},
  {0, 0.2, 0, 0, 0.2, 0, 0.1, 0.3},
  {0.3, 0, 0, 0.1, 0, 0.3, 0.1, 0.2},
  {0, 0, 0.1, 0, 0.1, 0, 0, 0},
  {0.1, 0, 0, 0.2, 0, 0.4, 0.1, 0},
  {0, 0.2, 0.2, 0.1, 0, 0, 0.1, 0},
  {0, 0, 0.3, 0, 0.1, 0.3, 0.2, 0},
  {.05, 0.2, .18, 0.2, 0.2, 0, 0, 0},
}


\def\matrixB{
  {0, 0.3, 0, 0.2, 0.3, 0, 0.1, 0},
  {0.2, 0, .24, 0, 0, 0.2, 0, 0},
  {\textcolor{red}{\textbf{0}}, 0.4, 0.1, 0.2, 0.1, 0, 0, 0},
  {0, 0.2, 0, 0, 0.2, 0, 0.1, 0.3},
  {0.3, 0, 0, 0.1, 0, 0.3, 0.1, 0.2},
  {0, 0, 0.1, 0, 0.1, 0, 0, 0},
  {\textcolor{red}{\textbf{0}}, 0, 0, 0.2, 0, 0.4, 0.1, 0},
  {0, 0.2, 0.2, 0.1, 0, 0, 0.1, 0},
  {0, 0, 0.3, 0, 0.1, 0.3, 0.2, 0},
  {\textcolor{red}{\textbf{0}}, 0.2, .18, 0.2, 0.2, 0, 0, 0},
}

\def\matrixC{
  {0, 0.3, 0, 0.2, 0.3, 0, 0.1, 0},
  {0.2, 0, .24, 0, 0, \textcolor{red}{\textbf{0}}, 0, 0},
  {\textcolor{red}{\textbf{0}}, 0.4, \textcolor{red}{\textbf{0}}, 0.2, \textcolor{red}{\textbf{0}}, 0, 0, 0},
  {0, \textcolor{red}{\textbf{0}}, 0, 0, 0.2, 0, \textcolor{red}{\textbf{0}}, 0.3},
  {0.3, 0, 0, \textcolor{red}{\textbf{0}}, 0, \textcolor{red}{\textbf{0}}, \textcolor{red}{\textbf{0}}, \textcolor{red}{\textbf{0}}},
  {0, 0, \textcolor{red}{\textbf{0}}, 0, \textcolor{red}{\textbf{0}}, 0, 0, 0},
  {\textcolor{red}{\textbf{0}}, 0, 0, \textcolor{red}{\textbf{0}}, 0, 0.4, \textcolor{red}{\textbf{0}}, 0},
  {0, \textcolor{red}{\textbf{0}}, 0.2, \textcolor{red}{\textbf{0}}, 0, 0, \textcolor{red}{\textbf{0}}, 0},
  {0, 0, 0.3, 0, \textcolor{red}{\textbf{0}}, 0.3, 0.2, 0},
  {\textcolor{red}{\textbf{0}}, \textcolor{red}{\textbf{0}}, \textcolor{red}{\textbf{0}}, 0.2, 0.2, 0, 0, 0},
}

\newcommand{\drawMatrix}[3]{
    \begin{scope}[xshift=#1 cm, yshift=#2 cm]
        \coordinate (topLeft) at (0.5, -0.5);
        \coordinate (bottomLeft) at (0.5, -10.5);
        \coordinate (topRight) at (8.5, -0.5);
        \coordinate (bottomRight) at (8.5, -10.5);

        \foreach \i [count=\row] in #3 {
          \foreach \x [count=\col] in \i {
            \node at (\col,-\row) {\x};
          }
        }

        \draw[thick] (topLeft) -- ++(-0.3,0) -- ++(0,-10) -- ++(0.3,0);
        \draw[thick] (topRight) -- ++(0.3,0) -- ++(0,-10) -- ++(-0.3,0);
  \end{scope}
}

\draw[thick, purple, dashed, rounded corners, fill=purple, fill opacity=0.2] (0.5, -3.5) rectangle (8.5, -2.5);
\draw[thick, blue, dashed, rounded corners, fill=blue, fill opacity=0.2] (0.5, -0.5) rectangle (1.5, -10.5);
\drawMatrix{0}{0}{\matrixA}
\draw[->, thick] (9.2, -5) -- (10, -5);

\draw[thick, green, dashed, rounded corners, fill=green, fill opacity=0.2] (11.5, -0.5) rectangle (12.5, -10.5);
\drawMatrix{11}{0}{\matrixB}
\node at (21, -5) {$\cdots$};

\draw[thick, purple, dashed, rounded corners, fill=purple, fill opacity=0.2] (22.5, -3.5) rectangle (30.5, -2.5);
\drawMatrix{22}{0}{\matrixC}

\end{tikzpicture}
}
\caption{Application of \gls{setsketch} with $\alpha=0.4$ to a toy set of sparse vectors $\mathcal{S}$ rendered as a matrix (leftmost): vectors become rows (\textcolor{purple}{purple box} is a vector); dimensions columns (\textcolor{blue}{blue box} is a dimension). In column $i$, the largest $\alpha$ portion of the nonzero entries are kept while the rest are reset. The center matrix shows that process for the first dimension. Rightmost is the final matrix $\tilde{\mathcal{S}}$ where each vector is sketched in the context of other vectors.}
\label{fig:set-sketch}
\end{figure}

We define the following notion first which is then used in the description of the algorithm. The algorithm is schematically described in Figure~\ref{fig:set-sketch}.
\begin{definition}
    \label{definition:dimension-density}
    Let $L_i = \{ j \;\mathit{s.t.}\; u^{(j)}_i \neq 0 \;\forall\; u^{(j)} \in \mathcal{S} \}$ be the list of vectors in the set $\mathcal{S}$ whose $i$-th coordinate is nonzero. We say that $\mathcal{S}$ is $\rho$-dense ($\rho \in [0, 1]$) in the $i$-th dimension with $\rho = \lvert L_i \rvert / \lvert \mathcal{S} \rvert$.
\end{definition}

\begin{definition}[Set \glsentrylong{sketch}]
\label{definition:global-alpha-mss}
Fix $\alpha \in (0, 1]$. Represent a set of sparse vectors $\mathcal{S} \subset \B$ in matrix form with vectors as rows and dimensions as columns. For each column $i$ with density $\rho_i$, keep the $\lambda_i = \lceil \alpha \rho_i \lvert \mathcal{S} \rvert \rceil$ largest values, and set all other values to $0$. The $j$-th row in the resulting matrix $\tilde{\mathcal{S}}$ is the \gls{setsketch} of the $j$-th vector.
\end{definition}

\subsubsection{Analysis of \gls{setsketch}}

In this section, we show that \gls{setsketch} has the property that its parameter $\alpha$ adapts to vectors based on their $\ell_1$ mass. To do that, we first establish that the \gls{setsketch} of a fixed vector $u$ in a randomly sampled set $\mathcal{S}$ is its \gls{sketch} in expectation.

\begin{theorem}
\label{theorem:expected-global-alpha-mss}
    In expectation over sets $\mathcal{S} \sim \mathcal{D}$, where dimensions are independent or weakly-dependent, the \gls{setsketch} $\tilde{u}$ of $u$ is the \gls{sketch} of $u$: $\E_{\mathcal{S} \sim \mathcal{D}}[\lVert \tilde{u} \rVert_1] = \alpha \lVert u \rVert_1$.
\end{theorem}
\begin{proof}
    Each $u_i$ of a vector $u$ is included in the \gls{setsketch} of $u$ if its rank among the $i$-th dimension of all vectors in $\mathcal{S}$ is less than $\lambda_i$: $r(u_i) \leq \lambda_i$. The probability of that event is equal to $\lambda_i/\rho_i \lvert \mathcal{S} \rvert = \alpha$. We have that:
    \begin{equation*}
        \E_{\mathcal{S} \sim \mathcal{D}}[\lVert \tilde{u} \rVert_1] = \sum_{i:\; u_i \neq 0} u_i \prob[\mathbbm{1}_{r(u_i) \leq \lambda_i}] =
        \sum_{i:\; u_i \neq 0} u_i \alpha = \alpha \lVert u \rVert_1.
    \end{equation*}
\end{proof}

The result above is in expectation. We are now interested in the deviation of $\lVert \tilde{u} \rVert_1$ from $\alpha \lVert u \rVert_1$. We present the following result to that end:
\begin{theorem}
    \label{theorem:deviation}
    Given the conditions of Theorem~\ref{theorem:expected-global-alpha-mss}, $\epsilon \in (0, 1)$, and a fixed $\mathcal{S}$, we have for each $u \in \mathcal{S}$ that:
    \begin{equation*}
        \mathbb{P}[\lVert \tilde{u} \rVert_1 - \alpha \lVert u \rVert_1 \geq \epsilon] \leq \exp\big( - \frac{\epsilon^2}{\alpha (1 - \alpha) \lVert u \rVert_2^2 + \frac{2}{3} \lVert u \rVert_1 \epsilon} \big).
    \end{equation*}
\end{theorem}
\begin{proof}
    Given the variance of $\lVert \tilde{u} \rVert_1$, and noting that $\lVert \tilde{u} \rVert_1 \in (0, \lVert u \rVert_1]$ is a bounded random variable, we can apply the one-sided Bernstein concentration inequality to derive the bound in the theorem. All that is left then is to derive the variance, which we do as follows. Letting $r(u_i)$ take the same definition as in the proof of Theorem~\ref{theorem:expected-global-alpha-mss}:
    \begin{equation*}
        \Var[\lVert \tilde{u} \rVert_1] = \sum_{i:\; u_i \neq 0} \Var[u_i \mathbbm{1}_{r(u_i) \leq \lambda_i}]
        = \sum u_i^2 \Var[\mathbbm{1}_{r(u_i) \leq \lambda_i}]
        = \lVert u \rVert_2^2 (\alpha - \alpha^2).
    \end{equation*}
    That completes the proof.
\end{proof}

It should be noted that, due to the independence assumption, the above results hold even if we restricted the space to an arbitrary subspace. In particular, we may restrict the space to the subspace spanned by the basis vectors corresponding to the nonzero coordinates of a query. Rewriting Theorem~\ref{theorem:expected-global-alpha-mss} gives that $\E_{\mathcal{S} \sim \mathcal{D}}[\lVert \tilde{u}_\mathcal{I} \rVert_1] = \alpha \lVert u_\mathcal{I} \rVert_1$, where $\mathcal{I}$ is the set of active dimensions in a query.

\bigskip
Let us examine what this bound implies. Suppose that $\alpha \ll 1$, so that $\alpha \lVert u \rVert_1 \ll \epsilon$. In the \msmarco dataset, for instance, the maximum $\ell_1$ norm is $0.7$, so that choosing $\alpha=0.1$ gives sketches whose $\ell_1$ norm is less than $0.07$. If $\lVert u \rVert_2 \ll 1$ (this value is less than $1e{-3}$ in \msmarco) then the first term in the denominator of the exponent vanishes, so that we are left with $\exp(-3\epsilon/2 \lVert u \rVert_1)$.

Now, fix $\epsilon$ and take vectors $u$ and $v$ such that $\lVert u \rVert_1 \ll \lVert v \rVert_1$. Then the upper-bound on the deviation is larger for $v$ than it is for $u$, implying that, with high probability, the \gls{setsketch} of $v$ results is a $\alpha_v$-\textsc{Mss} with $\alpha_v > \alpha$, while the \gls{setsketch} of $u$ results in an $\alpha_u$-MSS of $u$ with $\alpha_u \approx \alpha$.

That gives us the effect we sought to induce. To see that, suppose $\lVert u \rVert_1 \ll \lVert v \rVert_1$ implies with high probability that $\langle q, u \rangle < \langle q, v \rangle$---we have verified this earlier. Because $\alpha_u < \alpha_v$, more coordinates of $v$ are preserved in its sketch with high probability than coordinates of $u$. As such, according to Theorem~\ref{theorem:alpha-mss-bound}, the estimate of $\langle q, v \rangle$ is more accurate than $\langle q, u \rangle$, with high probability.

\glsreset{knn}

\section{\seismic: Our Sparse MIPS Algorithm} \label{sec:algorithm}
\begin{figure*}[t]
    \centering
    \includegraphics[width=\columnwidth]{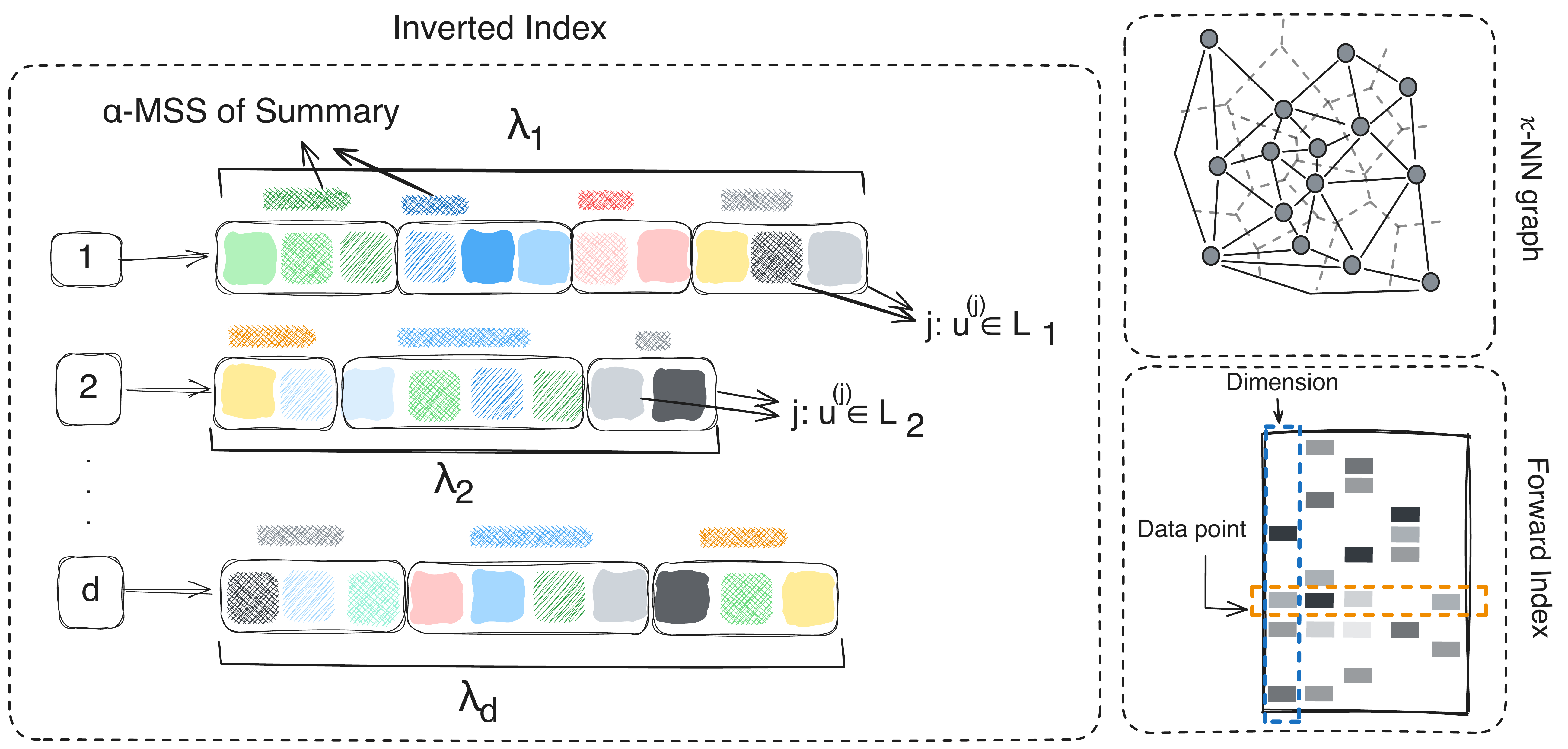}
    \caption{The design of \seismic. We apply \gls{setsketch} to the vector collection and build an inverted index. Each inverted list is then independently partitioned into geometrically-cohesive blocks, where a block is a set of data point identifiers along with an \gls{sketch} of a ``summary'' vector. The inner product of a query with the summary approximates the inner product attainable with the data points in that block. The forward index stores the complete vectors (i.e., all nonzero dimensions and coordinates). The \gls{knn}, which is stored as a table, records a mapping from a data point identifier to its $\kappa$ closest neighbors (by inner product).}
    \label{fig:inverted}
\end{figure*}

In this section, we describe our \gls{smips} algorithm dubbed \seismic. The algorithm relies on three data structures: an inverted index, a forward index, and (optionally) a \gls{knn}. In a nutshell, we use the inverted index (i.e., a mapping from dimensions to an inverted list: the id of data points whose coordinate in that dimension is nonzero) to identify a set of approximate nearest neighbors; the \gls{knn} (i.e., a graph where each node represents a data point, with edges that connect it to its $\kappa$ nearest neighbors) to refine and expand that approximate set; and the forward index (i.e., a mapping from data point id to its full vector representation) for exact inner product computation. Figure \ref{fig:inverted} gives an overview of the overall design.

\seismic is novel in the following ways. First, by applying \gls{setsketch} to a vector collection, it significantly reduces the size of the collection. Second, it partitions inverted lists into geometrically-cohesive blocks to reduce the number of data points for which it must compute exact inner product. Finally, it attaches a \emph{summary} to each block, whose inner product with a query approximates the inner product of the query with data points referenced in the block. We expand on these in detail in the following sections.

\subsection{Inverted Index}

\subsubsection{Sketching} \seismic heavily relies on the \gls{setsketch} algorithm---with a fixed $\alpha$, our first hyperparameter---as introduced in Section~\ref{section:sketch:global-mss}. The sketch was designed so that a small subset of a sparse vector's nonzero coordinates approximates its inner product with a query point with arbitrary accuracy, while preserving rankings of data points with high probability.

\subsubsection{Blocking of Inverted Lists} \label{sec:methodology:blocking}
\seismic also introduces a novel blocking strategy on inverted lists. It partitions the $i$-th inverted list, $L_i$, into $\beta \lvert L_i \rvert$ small blocks---our second hyperparameter. The rationale behind a blocked organization of an inverted list is to group together data points that are \emph{similar}, so as to facilitate a \emph{dynamic pruning} strategy, to be described shortly.

We defer the determination of similarity to a clustering algorithm. In other words, the data points referenced in the $i$-th inverted list are given as input to a clustering algorithm, which subsequently partitions them into $\beta \lvert L_i \rvert$ clusters. Each cluster is then turned into one block, consisting of the id of data points that belong to the same cluster. Conceptually, each block is ``atomic'' in the following sense: if the dynamic pruning algorithm decides we must visit a block, \emph{all} the data points referenced in that block are fully evaluated (i.e., their inner product with a query is computed exactly).

We note that any geometrical clustering algorithm may be readily used. We use a shallow variant~\cite{chierichetti2007clusterPruning} of K-Means as follows. Given an inverted list $L_i$, we uniformly-randomly sample $\beta \lvert L_i \rvert$ vectors from it, $\{ \mu^{(k)} \}_{k=1}^{\beta \lvert L_i \rvert}$, and use their corresponding data point as cluster representatives. For each $j \in L_i$, we find $k^\ast = \argmax_k \langle u^{(j)}, \mu^{(k)} \rangle$, and assign $j$ to the $k^\ast$-th cluster.

\subsubsection{Per-block Summary Vectors}
The query processing algorithm, to be presented later, requires an efficient way to determine if a block should be fully evaluated. To that end, \seismic leverages the concept of a \emph{summary} vector: a $d$-dimensional sparse vector that ``represents'' the data points referenced in a block. The summary vectors are stored in the inverted index, one per block, and are meant to serve as an efficient way to compute an approximation of the inner product between a query and the data points referenced in each block.

One realization of this idea is to upper-bound the full inner product attainable by data points referenced in a block. In other words, the $i$-th coordinate of the summary vector of a block would contain the maximum $u_i$ among the data points referenced in that block.

More precisely, our summary function $\phi: 2^\mathcal{S} \rightarrow [0, 1]^d$ takes a block $B$ from the universe of all blocks, and produces a vector whose $i$-th coordinate is simply:
\begin{equation}
    \label{equation:summary}
    \phi(B)_i = \max_{j \in B} u^{(j)}_i.
\end{equation}
This summary is \emph{conservative}: its inner product with the query is no less than the inner product between the query and any of the data points referenced in that block: $\langle q, \phi(B) \rangle \geq \langle q, u^{(j)} \rangle$ for all $j \in B$ and an arbitrary query $q$.

The caveat, however, is that the number of nonzero entries in summary vectors grows quickly with the block size.  That is the source of two potential issues: 1) the space required to store summaries increases; and 2) as inner product computation takes time proportional to the number of nonzero entries, the time to evaluate  a block could become unacceptably high.

We address that caveat by sketching the summary vectors using $\gamma$-\textsc{MSS}---$\gamma$ being our third hyperparameter---with the understanding that sketching takes away the conservatism property of the summary vectors. We further reduce the size of summaries by applying scalar quantization. With the goal of reserving a single byte for each value, we subtract the minimum value $m$ from each summary entry, and divide the resulting range into $256$ sub-intervals of equal size. A value in the summary is replaced with the index of the sub-interval it maps to. To reconstruct a value approximately, we multiply the id of its sub-interval by the size of the sub-intervals, then add $m$.

\subsection{Forward Index} As noted earlier, \seismic blends together three data structures. The first is an inverted index that tells us which data points to examine. To make it practical, we applied several layers of approximations that allow us to gain efficiency with a possible loss in accuracy. A forward index (built from the original vector collection) helps correct those errors and offers a way to compute the exact inner products between a query and the data points identified for full evaluation.

We must note that, data points referenced in the same block are not necessarily stored consecutively in the forward index. This is simply infeasible because the same data point may belong to different inverted lists and, thus, to different blocks. Because of this layout, computing the inner products may incur many cache misses, which are detrimental to query latency. In our implementation, we extensively use prefetching instructions to mitigate this effect.

\begin{algorithm}[t]
    \SetAlgoLined
    {\bf Input: }{Collection $\mathcal{S} \in \B$ of sparse vectors; $\alpha$: parameter of \gls{setsketch} to sketch $\mathcal{S}$; $\beta \in (0, 1)$: factor determining the number of blocks per inverted list; $\gamma$: parameter of $\gamma$-MSS for the summaries; $\kappa$: outgoing degree in the \gls{knn}.\\}
    \KwResult{\seismic index.}
    \begin{algorithmic}[1]
        \STATE $\mathcal{N}(u) \leftarrow \argmax^{(\kappa)}_{v \in \mathcal{S}} \; \langle u, v \rangle$ for all $u \in \mathcal{S}$ \label{algorithm:indexing:knn-graph} \triacomment{\small The \textmd{\gls{knn}}.}
        \STATE $\tilde{\mathcal{S}} \leftarrow $ \gls{setsketch} of $\mathcal{S}$ \label{algorithm:indexing:global-sketch}

        \FOR{$i \in \{ 1, \ldots, d\}$}
            \STATE $L_i \leftarrow \{ j \;|\; u^{(j)}_i \neq 0,\; u^{(j)} \in \tilde{\mathcal{S}} \}$ \label{algorithm:indexing:inverted-list} \triacomment{The inverted index}
            
            \STATE \textsc{Cluster} $L_i$ into $\beta \lvert L_i \rvert$ partitions, $\{ B_{i, j} \}_{j=1}^{\beta \lvert L_i \rvert}$ \label{algorithm:indexing:clustering} \triacomment{The blocks}
            \FOR{$1 \leq j \leq \beta$}
                \STATE $S_{i, j} \leftarrow$ $\gamma$-\textsc{MSS} of $\phi(B_{i, j})$ \triacomment{\small Sketched summaries using Equation~\eqref{equation:summary}} \label{algorithm:indexing:summary}
            \ENDFOR
        \ENDFOR
        \RETURN $\mathcal{N}, \{L_i\}_{i=1}^d$, $\{ B_{i, j}, S_{i, j} \} \; \forall_{1 \leq i \leq d,\; 1 \leq j \leq \beta \lvert L_i \rvert}$
    \end{algorithmic}
    \caption{Indexing with \seismic.}
    \label{algorithm:indexing}
\end{algorithm}

\subsection{\textsc{$\kappa$-nn} Graph} Our third data structure provides information that allows \seismic to \textit{refine} the candidate pool during query processing---we will describe that logic shortly. In particular, it constructs a graph from a collection of data points, wherein each node represents a data point with an outgoing edge to $\kappa$ other data points whose inner product with that node is maximal.

The resulting structure can be stored as a look-up table consisting of pairs of one data point id and a list of its $\kappa$ closest neighbors---our fourth hyperparameter. The \gls{knn} allows us to identify the nearest neighbors of a data point quickly. Formally, let us denote by $\mathcal{N}(u)$ the set of $\kappa$ closest data points to data point $u \in \mathcal{S}$:
\begin{equation}
    \mathcal{N}(u) = \argmax^{(\kappa)}_{v \in \mathcal{S}} \; \langle u, v \rangle.
    \label{equation:knn-graph}
\end{equation}
Setting $\kappa=0$ effectively removes the \gls{knn} from indexing and query processing altogether.

\subsection{Recap of Indexing} We summarize the discussion above in Algorithm \ref{algorithm:indexing}. When indexing a collection $\mathcal{S} \subset \B$, we first construct a \gls{knn} (Line~\ref{algorithm:indexing:knn-graph}). Next, we apply \gls{setsketch} to the collection (Line~\ref{algorithm:indexing:global-sketch}). For every dimension $i \in \{ 1, \dots, d\}$, we form its inverted list, recording only the data point identifiers (Line~\ref{algorithm:indexing:inverted-list}). Next, we apply clustering to each inverted list separately to derive at most $\beta \lvert L_i \rvert$ blocks (Line~\ref{algorithm:indexing:clustering}). Once data points are assigned to the blocks, we then build the $\gamma$-\textsc{MSS} of each block's summary (Line~\ref{algorithm:indexing:summary}).

\subsection{Query Processing} Algorithm \ref{algorithm:retrieval} shows the query processing logic in \seismic. We take an $\alpha_q$-MSS $\tilde{q}$ the query $q$ (Line~\ref{algorithm:retrieval:q-cut}). We then utilize our dynamic pruning strategy (Lines~\ref{algorithm:retrieval:summary-ip}--\ref{algorithm:retrieval:skip}) that allows the algorithm to skip blocks in the inverted lists corresponding to nonzero coordinates in $\tilde{q}$.

\begin{algorithm}[t]
    \SetAlgoLined {\bf Input: }{$q$: query; $k$: number of points to return; $\alpha_q$: parameter to $\alpha_q$-MSS to sketch the query; {\heapfactor}: a correction factor to rescale the summary inner product; $\mathcal{N}$, $L_i$'s and $(B_{i, j}, S_{i, j})$'s produced by Algorithm~\ref{algorithm:indexing}.}\\
    \KwResult{ A \heap\ with the top-$k$ data points.}
    
    \begin{algorithmic}[1]
        \STATE $\tilde{q} \leftarrow$ $\alpha_q$-MSS of $q$ \label{algorithm:retrieval:q-cut}
        \STATE \heap $\leftarrow \emptyset$
        \FORALL{$i \textit{ s.t. } \tilde{q}_i \neq 0$} \label{algorithm:retrieval:traversal}
            \STATE $r_j \leftarrow \langle q, S_{i, j} \rangle \quad \forall_{1 \leq j \leq \beta \lvert L_i \rvert}$ \label{algorithm:retrieval:summary-ip}
            \FORALL{$B_{i, j} \in L_i$ sorted in decreasing order by $r_j$}
                \IF{$\heap.{\sf len()} = k$ and $r_j < \frac{\heap.{\sf min()}}{\heapfactor}$}
                    \STATE {\bf continue} \triacomment{Skip the block}             \label{algorithm:retrieval:skip}
                \ENDIF
                \FOR{$d \in B_{i, j}$}
                    \STATE $p = \langle q, {\sf ForwardIndex}[d] \rangle$ \triacomment{Fully evaluate the data point with id $d$}
                    \IF{$\heap.{\sf len()} < k$ or $p > \heap.{\sf min()}$ }                 \STATE \heap.{\sf insert}$((p, d))$
                    \ENDIF
                    \IF{$\heap.{\sf len()} = k+1$}
                        \STATE \heap.{\sf pop\_min()}
                    \ENDIF
                \ENDFOR
            \ENDFOR
        \ENDFOR

        \STATE $S \leftarrow$ the ids of the vectors in the \heap \label{algorithm:retrieval:knn-graph:begin}
        \FOR{$j \in S$}
            \FOR{$v \in \mathcal{N}(u^{(j)})$}
                \STATE $p = \langle q, {\sf ForwardIndex}[v] \rangle$ \triacomment{Fully evaluate points in the expanded set $S_+$}
                \IF{$\heap.{\sf len()} < k$ or $p > \heap.{\sf min()}$ }
                    \STATE \heap.{\sf insert}$((p, v))$
                \ENDIF
                \IF{$\heap.{\sf len()} = k+1$}
                    \STATE \heap.{\sf pop\_min()} \label{algorithm:retrieval:knn-graph:end}
                \ENDIF
            \ENDFOR
        \ENDFOR
        
        \RETURN \heap
    \end{algorithmic}
    \caption{Query processing with \seismic.\label{algorithm:retrieval}}
\end{algorithm}

\seismic adopts a coordinate-at-a-time traversal (Line~\ref{algorithm:retrieval:traversal}) of the inverted index. For each nonzero coordinate $i$ of $\tilde{q}$, the algorithm sorts the blocks in $L_i$ by the inner product of their summary vectors with $q$ (referred to as ``summary score'') in descending order. Note that, inner products of $q$ with summaries in an inverted list can be computed efficiently with sparse matrix-vector multiplication. Sorting the blocks is also cheap as $\beta \lvert L_i \rvert$ is typically small.

The algorithm then visits the blocks in the sorted order. The data points referenced in a block are fully evaluated if the block's summary score is greater than a fraction (denoted by $\heapfactor \in (0, 1]$, the second hyperparameter) of the minimum inner product in the Min-\heap{}. That means the forward index retrieves the complete data point referenced in the selected block and computes their inner products with $q$. A data point whose inner product is greater than the minimum score in the Min-\heap{} is inserted into the heap.

Once the above concludes, we take the set of data points in the heap, denoted by $S$. As Lines~\ref{algorithm:retrieval:knn-graph:begin}--\ref{algorithm:retrieval:knn-graph:end} show, we then form the \emph{expanded set} $S_+ = \bigcup_{u \in S} \big( \{ u \} \cup \mathcal{N}(u) \big)$, compute scores for data points in $S_+$, and return the top-$k$ subset. As noted previously, $\kappa=0$ disables the expansion of $S$.


\section{Experiments} \label{section:experiments}
We now evaluate \seismic{} experimentally. Specifically, we are interested in investigating the performance of \seismic along the following axes: accuracy, latency, space usage, indexing time, and scalability.

We begin by introducing our setup, including datasets and embedding models. We then evaluate \seismic with $\kappa=0$, so that the \gls{knn} is disabled. Next, we enable the \gls{knn} and study its impact on the evaluation axes above. We conclude our empirical evaluation by scaling up the collection of vectors by an order of magnitude.

\subsection{Setup}

\subsubsection{Datasets}
We experiment on three publicly-available datasets. These include:
\begin{itemize}
    \item \msmarco~\cite{nguyen2016msmarco}: a collection of $8.8$M text passages in English with a set of $6{,}980$ queries;
    \item Natural Questions (\nq{}): a \beir~\cite{thakur2021beir} collection of $2.68$M questions and $7{,}842$ queries; and,
    \item \msmarcodos: a collection of $138$ million passages with $3{,}903$ queries.
\end{itemize}
We use the last dataset in our scalability study only.

\subsubsection{Sparse embedding models}
We evaluate all \gls{smips} algorithms on the embeddings of the above datasets using four embedding models:
\begin{itemize}
\item \splade~\cite{formal2022splade}. Each nonzero entry is the importance weight of a term in the BERT~\cite{devlin2019bert} WordPiece~\cite{wordpiece} vocabulary consisting of $30$,$000$ terms. We use the \texttt{cocondenser-ensembledistil}\footnote{\url{https://huggingface.co/naver/splade-cocondenser-ensembledistil}} version of \splade. The number of nonzero entries in data points (query points) is, on average, $119$ ($43$) for \msmarco; $153$ ($51$) for \nq; and, $127$ ($44$) for \msmarcodos.

\item Efficient \splade~\cite{lassance2022efficient-splade}. Similar to \splade, but whose embeddings typically have a larger number of nonzero coordinates in data points and a smaller number in query points. For example, there are $181$ ($5.9$) nonzero entries in \msmarco data (query) points. We use the \texttt{efficient-splade-V-large}\footnote{Checkpoints at \url{https://huggingface.co/naver/efficient-splade-V-large-doc} and \url{https://huggingface.co/naver/efficient-splade-V-large-query}.}. We refer to this model as \esplade{}.

\item \spladeT~\cite{lassance2024spladev3}. An improved variant of \splade that incorporates a modified objective and distillation strategy. The number of nonzero entries in data (query) points is, on average, $168$ ($24$) for \msmarco.

\item \unicoil~\cite{unicoil,ma2022document}. Expands a text passge with relevant terms generated by DocT5Query~\cite{nogueira2019doc2query}.
There are, on average, $68$ ($6$) nonzero entries in \msmarco data (query) points.
\end{itemize}
We note that, we generate the \splade and \esplade embeddings using the original code published on GitHub.\footnote{\url{https://github.com/naver/splade}}

\subsubsection{Baselines}
We compare \seismic with six state-of-the-art \gls{smips} algorithms. Two of these are the winning solutions of the ``Sparse Track'' at the 2023 BigANN Challenge\footnote{\url{https://big-ann-benchmarks.com/neurips23.html}} at NeurIPS:
\begin{itemize}
    \item \grassrma: A graph-based method for dense vectors adapted to sparse vectors that appears in the BigANN challenge as ``\textsc{sHnsw}.''\footnote{C++ code is publicly available at \url{https://github.com/Leslie-Chung/GrassRMA}.}
    \item \pyann: Another graph-based ANN solution.\footnote{C++ code is publicly available at \url{https://github.com/veaaaab/pyanns}.}
\end{itemize}
The other baselines are as follows:
\begin{itemize}
    \item \ioqp~\cite{mpg22-desires}: Impact-sorted query processor written in Rust. We choose \ioqp because it is known to outperform JASS~\cite{jass2015}, another widely-adopted impact-sorted query processor.
    \item \bruchetal~\cite{bruch2023bridging}: An inverted index where lists are partitioned into blocks through clustering. At query time, after finding the $N$ closest clusters to the query, a coordinate-at-a-time algorithm traverses the inverted lists. The solution is approximate because only the data points that belong to top $N$ clusters are considered.
    \item \pisa~\cite{MSMS2019}: An inverted index-based library based on \texttt{ds2i}~\cite{pefi-SIGIR14} that uses highly-optimized blocked variants of WAND. \pisa is \emph{exact} as it traverses inverted lists in a rank-safe manner.
    \item \kannolo~\cite{kannolo}: a new implementation of HNSW~\cite{hnsw2020} written in Rust for approximate nearest neighbors search, targeting both dense and sparse domains.\footnote{Code obtained from \url{https://github.com/TusKANNy/kannolo}} We choose this solution because, as Delfino \emph{et al.} show, it outperforms both \grassrma and \pyann in terms of search time.
\end{itemize}

We also considered the method by Lassance \etal~\cite{lassance2023static-pruning}. Their approach statically prunes either inverted lists (by keeping $p$-quantile of elements), data points (by keeping a fixed number of top entries), or all coordinates whose value is  below a threshold. While simple, their method is only able to speed up query processing by 2--4$\times$ over \pisa on \esplade embeddings of \msmarco. We found it to be ineffective on \splade and generally far slower than \grassrma and \pyann. As such we do not include it in our discussion.

\subsubsection{Evaluation Metrics}
We evaluate all methods using three metrics:
\begin{itemize}
    \item Latency ($\mu$sec.). The time elapsed, in \emph{microseconds}, from the moment a query vector is presented to the index to the moment it returns the approximate set of $k$ nearest data vectors running in single thread mode. Latency does not include embedding time.
    \item Accuracy. The percentage of true nearest neighbors recalled in the returned set, per Definition~\ref{definition:ann-accuracy}. By measuring the recall of an approximate set given the exact top-$k$ set, we study the impact of the different levers in an algorithm on its overall accuracy as a retrieval engine.
    \item Index size (MiB). The space the index occupies in memory.
\end{itemize}

\subsubsection{Reproducibility and Hardware Details}
We implemented \seismic{} in Rust.\footnote{Our code is publicly available at \url{https://github.com/TusKANNy/seismic}.} We compile \seismic{} by using the version $1{.}77$ of Rust and use the highest level of optimization made available by the compiler. We conduct experiments on a server equipped with one Intel i9-9900K CPU with a clock rate of $3{.}60$ GHz and $64$ GiB of RAM. The CPU has $8$ physical cores and $8$ hyper-threaded ones. We query the index using a single thread.

\subsection{\seismic without the \textsc{$\kappa$-nn} Graph}
We now present our experimental results for a configuration of \seismic where $\kappa=0$, effectively removing the \gls{knn} from indexing and search subroutines.

\subsubsection{Hyperparameter tuning}
We build \ioqp and \pisa indexes using Anserini\footnote{\url{https://github.com/castorini/anserini}} and apply recursive graph bisection~\cite{mpm21-sigir}. For \ioqp, we vary the \emph{fraction} (of the total collection) hyperparameter in $[0.1, 1]$ with step size of $0.05$. For \bruchetal, we sketch data points using \sinnamon and a sketch size of $1{,}024$, and build $4 \sqrt{N}$ clusters, where $N$ is the number of data points in the collection. For \grassrma and \pyann, we build different indexes by running all possible combinations of $ef_c \in  \{1000, 2000\}$ and $M \in \{16, 32, 64, 128, 256 \}$. During search we test $ef_s \in [5, 100]$ with step size $5$, then $[100, 400]$ with step $10$, $[100, 1000]$ with step $100$, and finally $[1000, 5000]$ with step $500$. We apply early stopping when accuracy saturates.

Our grid search for \seismic on \msmarco is over: $\alpha \in \{0.05, 0.1, 0.15, 0.2, 0.3, 0.4, 0.5\}$; $\beta \in \{0.1, 0.2, 0.3\}$ with step $x$, and $\gamma \in [0.1, 0.5]$ with step size $0.1$.  Best results on \splade are achieved when $\alpha=0.1$, $\beta=0.3$, and $\gamma=0.4$. Best results with \esplade with $\alpha=0.1$, $\beta=0.2$, and $\gamma=0.4$. Best results for \unicoil with $\alpha=0.2$, $\beta=0.2$, and $\gamma=0.4$.
The grid search for \seismic on \nq is over: 
$\alpha \in \{0.05, 0.1, 0.15, 0.2, \}$; $\beta \in \{0.1, 0.2, 0.3\}$, and $\gamma \in \{0.4, 0.5, 0.6 \}$. Best results are achieved when $\alpha=0.1$, $\beta=0.3$, and $\gamma=0.5$. \seismic uses 8-bit scalar quantization for summaries, and matrix multiplication to efficiently compute the product of a query vector with all quantized summaries of an inverted list.

\begin{table*}[t]
\caption{Mean latency ($\mu$sec/query) at different accuracy$@10$ cutoffs with speedup (in parenthesis) gained by \seismic over others.\label{table:seismic-without-knn}}
	\centering
    \adjustbox{max width=\textwidth}{%
       	\begin{tabular}{lr@{\hspace{.75\tabcolsep}}rr@{\hspace{.75\tabcolsep}}rr@{\hspace{.75\tabcolsep}}rr@{\hspace{.75\tabcolsep}}rr@{\hspace{.75\tabcolsep}}rr@{\hspace{.75\tabcolsep}}r}

        \toprule
        \multicolumn{13}{c}{\splade on \msmarco} \\
        \midrule
        Accuracy (\%) & \multicolumn{2}{c}{90} & \multicolumn{2}{c}{91} & \multicolumn{2}{c}{92} & \multicolumn{2}{c}{93} & \multicolumn{2}{c}{94} & \multicolumn{2}{c}{95} \\
        \midrule
       \ioqp & $17$,$423$ & (96.8$\times$) & $17$,$423$ & (96.8$\times$) & $18$,$808$ & (90.0$\times$) & $21$,$910$ & (104.8$\times$) & $24$,$382$ & (101.6$\times$) & $31$,$843$ & (98.3$\times$) \\
        \bruchetal
        &$4{,}169$ & (23.2$\times$) & $4{,}984$ & (27.7$\times$) & $6{,}442$ & (30.8$\times$) & $7{,}176$ & (34.3$\times$) & $8{,}516$ & (35.5$\times$) & $10{,}254$ & (31.6$\times$)
        \\
        \grassrma & $807$ & (4.5$\times$) & $867$ & (4.8$\times$) & $956$ & (4.6$\times$) & $1{,}060$ & (5.1$\times$) & $1{,}168$ & (4.9$\times$) & $1{,}271$ & (3.9$\times$) \\
        \pyann & $489$ & (2.7$\times$) & $539$ & (3.0$\times$) & $603$ & (2.9$\times$) & $654$ & (3.1$\times$) & $845$ & (3.5$\times$) & $1{,}016$ & (3.1$\times$) \\
        \kannolo & 537 & (3.0$\times$) &586 & (3.3$\times$) & 629 & (3.0$\times$) & 673 & (3.2$\times$) & 743 & (3.1$\times$) & 836 & (2.6$\times$) \\
        
        \textbf{\seismic ($\kappa=0$)} & 180 & \multicolumn{1}{c}{--} & 180 & \multicolumn{1}{c}{--} & 209 & \multicolumn{1}{c}{--} & 209 & \multicolumn{1}{c}{--} & 240 & \multicolumn{1}{c}{--} & 324 & \multicolumn{1}{c}{--} \\
       
        \toprule
        \multicolumn{13}{c}{\esplade on \msmarco} \\
        \midrule
        \ioqp & $7{,}857$ & (45.9$\times$) & $8{,}382$ & (46.8$\times$) & $8{,}892$ & (49.7$\times$) & $9{,}858$ & (51.3$\times$) & $10{,}591$ & (35.9$\times$) & $11{,}536$ & (37.8$\times$) \\
        \bruchetal & $4{,}643$ & (27.2$\times$) & $5{,}058$ & (28.3$\times$) & $5{,}869$ & (32.8$\times$) & $6{,}599$ & (34.4$\times$) & $7{,}555$ & (25.6$\times$) & $8{,}962$ & (29.4$\times$) \\
        \grassrma & $2{,}074$ & (12.1$\times$) & $2{,}658$ & (14.8$\times$) & $2{,}876$ & (16.1$\times$) & $3{,}490$ & (18.2$\times$) & $4{,}431$ & (15.0$\times$) & $5{,}141$ & (16.9$\times$) \\
        \pyann & $1{,}685$ & (9.9$\times$) & $1{,}702$ & (9.5$\times$) & $2{,}045$ & (11.4$\times$) & $2{,}409$ & (12.5$\times$) & $3{,}119$ & (10.6$\times$) & $4{,}522$ & (14.8$\times$) \\
        \kannolo & $1{,}310$ & (7.7$\times$) & $1{,}432$ & (8.0$\times$) & $1{,}649$ & (9.2$\times$) & $1{,}864$ & (9.7$\times$) & $2{,}369$ & (8.0$\times$) & $2{,}864$ & (9.4$\times$) \\
        
        \textbf{\seismic ($\kappa = 0$)} & $171$ & \multicolumn{1}{c}{--} & $179$ & \multicolumn{1}{c}{--} & $179$ & \multicolumn{1}{c}{--} & $192$ & \multicolumn{1}{c}{--} & $295$ & \multicolumn{1}{c}{--} & $305$ & \multicolumn{1}{c}{--}\\

    \toprule
    \multicolumn{13}{c}{\unicoil on \msmarco} \\
    \midrule
    \ioqp & 
    $22{,}278$ & (210.2$\times$) & $25{,}060$ & (217.9$\times$) & $26{,}541$ & (215.8$\times$) & $30{,}410$ & (230.4$\times$) & $33{,}327$ & (173.6$\times$) & $34{,}061$ & (168.6$\times$) \\
    \bruchetal & $6{,}375$ & (60.1$\times$) & $7{,}072$ & (61.5$\times$) & $8{,}192$ & (66.6$\times$) & $9{,}207$ & (69.8$\times$) & $10{,}306$ & (53.7$\times$) & $12{,}308$ & (60.9$\times$) \\
    \grassrma & 
    $1{,}318$ & (12.4$\times$) & $1{,}434$ & (12.5$\times$) & $1{,}812$ & (14.7$\times$) & $2{,}004$ & (15.2$\times$) & $2{,}168$ & (11.3$\times$) & $2{,}668$ & (13.2$\times$) \\
    \pyann & $1{,}133$ & (10.7$\times$) & $1{,}456$ & (12.7$\times$) & $1{,}741$ & (14.2$\times$) & $1{,}755$ & (13.3$\times$) & $2{,}061$ & (10.7$\times$) & $2{,}973$ & (14.7$\times$) \\
    \kannolo & $1{,}085$ & (10.2$\times$) &  $1{,}188$ & (10.3$\times$) & $1{,}301$ & (10.6$\times$) & $1{,}468$ & (11.1$\times$) & $1{,}691$ & (8.8$\times$) & $1{,}961$ & (9.7$\times$) \\
    
    \textbf{\seismic ($\kappa=0$)} & 106 & \multicolumn{1}{c}{--} & 115 & \multicolumn{1}{c}{--} & 123 & \multicolumn{1}{c}{--} & 132 & \multicolumn{1}{c}{--} & 192 & \multicolumn{1}{c}{--} & 202 & \multicolumn{1}{c}{--} \\
    \bottomrule
        
    \toprule
    \multicolumn{13}{c}{\splade on \nq} \\
    \midrule
    \ioqp & $8{,}313$ & (47.2$\times$) & $8{,}854$ & (43.0$\times$) & $9{,}334$ & (38.3$\times$) & $11{,}049$ & (45.3$\times$) & $11{,}996$ & (42.2$\times$) & $14{,}180$ & (42.8$\times$) \\
    \bruchetal & $3{,}862$ & (21.9$\times$) & $4{,}309$ & (20.9$\times$) & $4{,}679$ & (19.2$\times$) & $5{,}464$ & (22.4$\times$) & $6{,}113$ & (21.5$\times$) & $6{,}675$ & (20.2$\times$)\\
    \grassrma  & $1{,}000$ & (5.7$\times$) & $1{,}138$ & (5.5$\times$) & $1{,}208$ & (5.0$\times$) & $1{,}413$ & (5.8$\times$) & $1{,}549$ & (5.5$\times$) & $2{,}091$ & (6.3$\times$) \\
    \pyann & $610$ & (3.5$\times$) & $668$ & (3.2$\times$) & $748$ & (3.1$\times$) & $870$ & (3.6$\times$) & $1{,}224$ & (4.3$\times$) & $1{,}245$ & (3.8$\times$) \\
    \kannolo & 471 & (2.7$\times$) & 521 & (2.5$\times$) & 567 & (2.3$\times$) &  606 & (2.5$\times$) & 702 & (2.5$\times$) & 805 & (2.4$\times$) \\
    
    \textbf{\seismic ($\kappa=0$)} & 176 & \multicolumn{1}{c}{--} & 206 & \multicolumn{1}{c}{--} & 244 & \multicolumn{1}{c}{--} & 244 & \multicolumn{1}{c}{--} & 284  & \multicolumn{1}{c}{--} & 331  & \multicolumn{1}{c}{--} \\
    \bottomrule

    \end{tabular}}
\end{table*}

\subsubsection{Accuracy-Latency Trade-off}
Table~\ref{table:seismic-without-knn} details \gls{anns} performance in terms of average per-query latency for \splade, \esplade, and \unicoil on \msmarco, and \splade on \nq. We frame the results as the trade-off between effectiveness and efficiency. In other words, we report mean per-query latency at a given accuracy level.

The results on these datasets show \seismic's remarkable relative efficiency, reaching a latency that is often one to two orders of magnitude smaller. Overall, \seismic consistently outperforms all baselines at all accuracy levels, including \grassrma and \pyann, which in turn perform better than other inverted index-based baselines---confirming the findings of the BigANN Challenge.

We make a few additional observations. \ioqp appears to be the slowest method across datasets. This is not surprising considering \ioqp is not designed with the distributional properties of sparse embeddings in mind. \bruchetal generally improves over \ioqp, but \seismic speeds up query processing further. In fact, the minimum speedup over \ioqp (\bruchetal) on \msmarco is $84.6\times$ ($22.3\times$) on \splade, $30.7\times$ ($20.9\times$) on \esplade, and $181\times$ ($54.8\times$) on \unicoil.

\seismic consistently outperforms \grassrma and \pyann by a substantial margin, ranging from $2.6\times$ (\splade on \msmarco) to $21.6\times$ (\esplade on \msmarco) depending on the level of accuracy. In fact, as accuracy increases, the latency gap between \seismic and the two graph-based methods widens. This gap is much larger when query vectors are sparser, such as with \esplade embeddings. That is because, when queries are highly sparse, inner products between queries and data points become smaller, reducing the efficacy of a greedy graph traversal. As one statistic, \pyann over \esplade embeddings of \msmarco visits roughly $40{,}000$ data points to reach $97\%$ accuracy, whereas \seismic evaluates just $2{,}198$ data points.

We highlight that \pisa is the slowest (albeit, \emph{exact}) solution. On \msmarco, \pisa processes queries in about $100{,}325$ microseconds on \splade embeddings. On \esplade and \unicoil, its average latency is $7{,}947$ and $9{,}214$ microseconds, respectively. We note that its high latency on \splade is largely due to the much larger number of nonzero entries in queries.

Finally, we must note that we have conducted a comparison of \seismic with another recent algorithm named BMP~\cite{bmp} for completeness. While BMP performs similarly to \seismic in terms of the index size and accuracy, it has a considerably larger per-query latency. As one data point, setting the memory budget to $8$GB and performing search over the \splade embeddings of \msmarco, BMP executes each query in $3.2$ milliseconds on average while \seismic takes $0.7$ milliseconds per query to reach the same level of accuracy.

\subsubsection{Space and Build Time}
Table~\ref{table:results-rq2} records the time it takes to index the entire \msmarco collection embedded with \splade with different methods, and the size of the resulting index. We left out the build time for \ioqp because its index construction has many external dependencies (such as Anserini and graph bisection) that makes giving an accurate estimate difficult. We perform this experiment on a machine with two Intel Xeon Silver 4314 CPUs clocked at 2.40GHz, with 32 physical cores plus 32 hyper-threaded ones and 512 GiB of RAM. We build the indexes by using multi-threading parallelism with $64$ cores.

Trends for other datasets are similar to those reported in Table~\ref{table:results-rq2}. Notably, indexes produced by approximate methods are larger. That makes sense: using more auxiliary statistics helps narrow the search space dynamically and quickly. Among the highly efficient methods, the size of \seismic's index is mild, especially compared with \grassrma. Importantly, \seismic builds its index in a fraction of the time it takes \pyann or \grassrma to index the collection.

\begin{table}[t]
    \caption{Index size and build time.\label{table:results-rq2}}
	\centering
\adjustbox{max width=\textwidth}{%
	\begin{tabular}{lrr}
    \toprule
    \multicolumn{3}{c}{\splade on \msmarco} \\
    \midrule
    Model & Index size (MiB) & Index build time (min.) \\
    \midrule
    \ioqp & $2{,}195$ & - \\
    \bruchetal & $8$,$830$ & $44$\\
    \grassrma & $10$,$489$ &$267$ \\
    \pyann & $5$,$262$ & $137$\\
    \kannolo & $7$,$369$& $147$ \\
    \textbf{\seismic ($\kappa=0$)} & $6$,$598$ &  $2$\\
    \bottomrule
	\end{tabular}}
\end{table}

\subsection{\seismic with the \textsc{$\kappa$-nn} Graph}
Now that we have compared the performance of \seismic against all baseline algorithms, we move to present experimental results for a configuration of \seismic \emph{with} the \gls{knn} (i.e., $\kappa > 0$). As we will see, including the \gls{knn} leads to better accuracy and faster search, but at the expense of increased index construction time and size.

We note that, because in the preceding section we concluded that \seismic is far faster and more accurate than baseline \gls{anns} algorithms, we limit our comparison in this section to different flavors of \seismic: one without the \gls{knn}, and another with.

Before we present our results, let us elaborate on how we construct the \gls{knn}. Constructing the (exact) graph is expensive due to its quadratic time complexity. As such, we relax the construction to an approximate (but almost-exact) graph: we connect each data point to its approximate set of top-$\kappa$ points. To find the $\kappa$ approximate set, we use \seismic (without the \gls{knn}). Storing the graph takes $ (\lfloor \log_{2} (N-1)\rfloor  + 1)N\kappa$ bits, where $N$ is the size of the collection.

\subsubsection{Hyperparameter tuning}
When building the two \seismic indexes, we first fix a memory budget as a multiple of the size of the forward index. We then sweep the hyperparameters as follows to find the best configuration that results in an index no larger than the budget:  $\alpha \in \{0.05, 0.1, 0.15, 0.2, 0.3, 0.4, 0.5\}$; $\beta \in \{0.1, 0.2, 0.3\}$, and $\gamma \in [0.1, 0.6]$ with step size $0.1$ and when the \gls{knn} is enabled, $\kappa \in \{ 10, 20, 30, 40, 50\}$. We set $\alpha_q \in [0.1, 0.9]$ with step size $0.1$ and $ \heapfactor \in \{0.7, 0.8, 0.9, 1.0 \}$, and report the best configuration.

\begin{table*}[t]
	\centering
    \adjustbox{max width=\textwidth}{%
   	\begin{tabular}{cccllllllllll}
        \toprule
         Embeddings & Budget & Accuracy (\%) & 90 & 91 & 92 & 93 & 94 & 95 & 96 & 97 & 98 & 99 \\
        \midrule
        \multirow{4}{*}{\splade} &
        \multirow{2}{*}{$1.5 \times$} &
        
        \seismic ($\kappa = 0$) & $183$ ($1.5\times$) & $207$ ($1.5\times$) & $236$ ($1.5\times$) & $313$ ($2.0\times$) & $348$ ($1.8\times$) & $517$ ($2.3\times$) & & $-$ & $-$ & $-$ \\
        & & \seismic ($\kappa > 0$) & 126 & 136 & 154 & 154 & 196 & 226 & 333 & 539 & -- & -- \\
        \cmidrule{2-13}
        &\multirow{2}{*}{$2 \times$}
        & \seismic ($\kappa=0$)
        & $174$ ($1.4\times$) & $174$ ($1.3\times$) & $209$ ($1.6\times$) & $209$ ($1.5\times$) & $249$ ($1.6\times$) & $255$ ($1.6\times$) & $301$ ($1.8\times$) & $349$ ($1.7\times$) & $480$ ($2.0\times$) & $715$ ($1.8\times$) \\
        & & \seismic ($\kappa>0$) & $ 121 $ & $ 134 $ & $ 134 $ & $ 143 $ & $156  $ & $156  $ & $168$ & $203$ & $245$ & $387$ \\
        \midrule
        \multirow{4}{*}{\spladeT} &
        \multirow{2}{*}{$1.5 \times$} &
        
        \seismic ($\kappa=0$) &
        $298$ ($1.1\times$) & $298 $ ($1.1\times$) & $298 $ ($1.1\times$) & $298 $ ($1.1\times$) & $ 298$ ($1.1\times$) & $438$ ($1.6\times$) & $438 $ ($1.6\times$) &  & -- & -- \\

        & & \seismic ($\kappa>0$) & $ 281 $ & $ 281 $ & $ 281 $ & $ 281 $ & $ 281 $ & $ 281 $ & $ 281 $ & $ 281 $ & $424$ & -- \\
        \cmidrule{2-13}
        & \multirow{2}{*}{$2 \times$}&
        \seismic ($\kappa=0$) &
        $152 $ ($0.6\times$) & $152 $ ($0.6\times$) & $176 $ ($0.7\times$) & $176 $ ($0.7\times$) & $ 201$ ($0.8\times$) & $ 222 $ ($0.9\times$) & $254 $ ($1.0\times$) & $ 288$ ($1.1\times$) & $ 328 $ ($1.2\times$) & $ 567 $ ($1.5\times$)
        \\
        & & \seismic ($\kappa>0$) & $ 253 $ & $ 253 $ & $ 253 $ & $ 253 $ & $ 253 $ & $ 253 $ & $ 253 $ & $ 253 $ & $266 $ & $ 371$ \\
        \bottomrule
    \end{tabular}}
	\caption{Mean latency ($\mu$sec/query) at different accuracy$@10$ cutoffs with speedup (in parenthesis) gained by adding the \gls{knn} on \msmarco. The ``Budget'' column indicates the memory budget as a multiple of the size of the forward index.\label{table:seismic-with-knn}}
    \vspace{-5mm}
\end{table*}

\subsubsection{Results}
Table~\ref{table:seismic-with-knn} compares the two configurations of \seismic on \msmarco, vectorized using two of the embedding models. In these experiments, we consider two memory budgets expressed as multiples of the size of the collection ($4$GB for \splade and $5{.}6$GB for \spladeT). We choose the best index configuration that respects the budget and report the fastest configuration reaching the accuracy cutoffs from $90\%$ to $99\%$.  We also report the speedup in parenthesis gained by adding the \gls{knn}.

It is clear that adding the \gls{knn} to \seismic leads to faster search across accuracy cutoffs. On \splade, this speedup is $2.2\times$ with a budget of $1.5\times$, and is up to $1.7\times$ with a $2.0\times$ memory budget. The same trend can be observed on \spladeT. Note that, because \spladeT embeddings are less sparse, the memory impact of storing the \gls{knn} is smaller.

Interestingly, the ``exact'' algorithm \pisa~\cite{MSMS2019} caps at $99\%$ accuracy due to quantization, which enables significant memory savings. At $99\%$ cutoff, \seismic (with \gls{knn}) takes $409 \mu s$, which is \emph{two orders of magnitude} faster than \pisa's $95{,}818 \mu s$ per query.

\subsection{Scalability of \seismic}
In this section, we apply \kannolo and \seismic to \msmarcodos, a massive dataset of $138$ million embedding vectors, and compare their performance. As before, let us explain our protocol for tuning the hyperparameters of the algorithm before presenting experimental results.

\subsubsection{Hyperparameter tuning}
As in the preceding section, we allocate memory budgets for the index as multiples of the dataset size, which is about $66$GB with $16$-bit floats for values and $16$-bit integers for components. We only consider hyperparameters that result in indexes that are up to $1.5\times$ the dataset size in one scenario, and $2\times$ in another.

\kannolo has the following hyperparameters: $M$ (number of neighbors per node), and $ef_c$ (number of nodes scanned to build a node's neighborhood). We sweep $M \in \{16, 32, 64 \}$ and $ef_c = 500$. For \seismic, we set: $\alpha \in {0.1, 0.2, 0.3}$ $\beta = 0.4$, and $\gamma=0.4$. We let $\kappa$ take values in $\{10, 20\}$.

Given an index, we process queries using the following hyperparameters and evaluate each configuration separately.
For \seismic, we vary $\alpha_q$ \ in $[0.1, 0.9]$, with step $0.1$, \heapfactor in $\{ 0.6, .. , 1.0\}$. For \kannolo, we vary $ef_s$ in $[10, 50]$ with step $5$, $[50, 100]$ with step $10$, $[100, 1500]$ with step $100$. For both methods, we consider the shortest time it takes the algorithm to reach top-$10$ accuracy of $\{90\%, 91\%, \ldots, 99\%\}$.

\subsubsection{Hardware details}
Due to a much larger collection size, we use a different machine for experiments in this section. We run experiments on a NUMA server with $1$TiB of RAM and four Intel Xeon Gold 6252N CPUs ($2{.}30$ GHz), with a total of $192$ cores ($96$ physical and $96$ hyper-threaded). Using the \texttt{numactl}\footnote{\url{https://linux.die.net/man/8/numactl}} tool, we enforce that the execution of the retrieval algorithms is done on a single CPU and its local memory. In this way, we avoid performance degradation due to non-uniform memory accesses across different CPUs. We note that, each CPU node has enough local memory ($256$GiB) to store entire indexes.

\subsubsection{Accuracy and latency}
\begin{figure}[!t]
    \centering
    \includegraphics[width=0.85\linewidth]{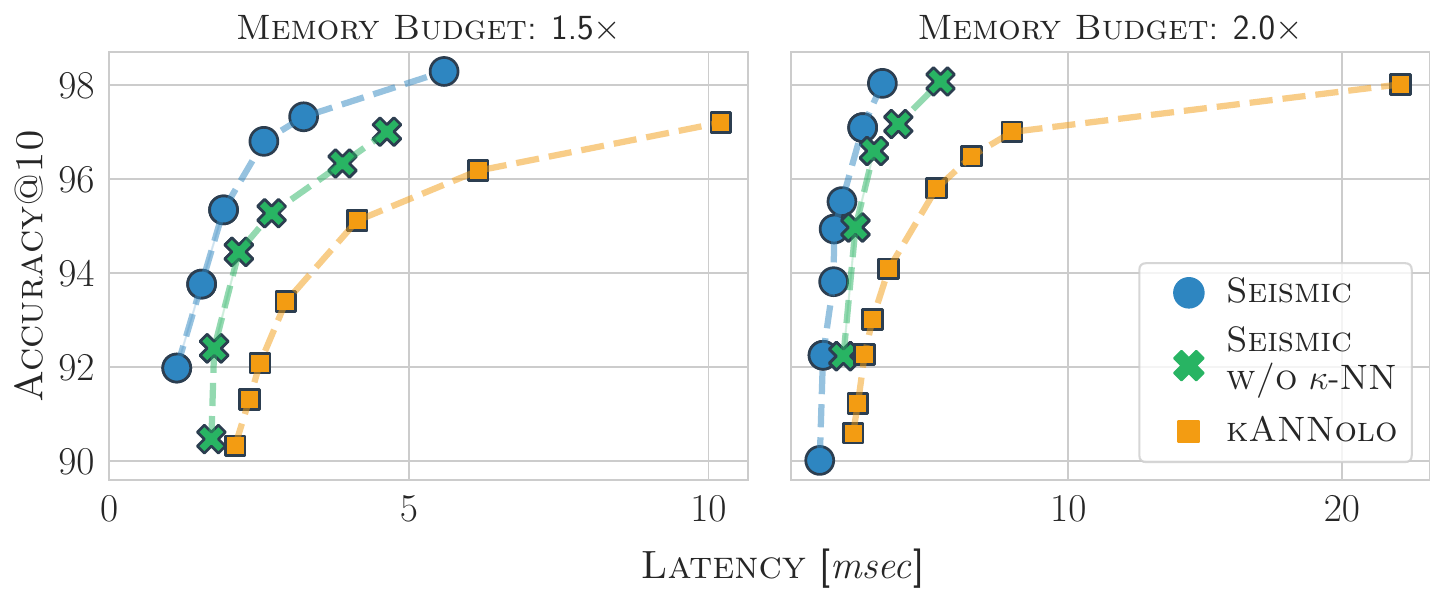}
    \caption{Comparison of \kannolo and \seismic (with and without \gls{knn}) by accuracy$@10$ as a function of query latency. We allow hyperparameters that result in an index whose size is at most $1.5\times$ (left) or $2\times$ (right) the size of the dataset. Note that, unlike previous figures, latency is measured in \emph{milliseconds}.}
    \label{fig:k10}
\end{figure}

Our first experiments, visualized in Figure~\ref{fig:k10}, compare \kannolo and \seismic by accuracy as a function of query latency for top-$10$ \gls{anns}. \seismic remains Pareto-efficient relative to \kannolo in both memory budget settings.

Interestingly, when we allow the algorithms to use more memory (i.e., $2\times$ the dataset size), the gap between \seismic and \kannolo widens as accuracy increases. That is thanks to the \gls{knn}, which gives an advantage in high accuracy scenarios ($>=97\%$), as validated by a comparison of \seismic with and without the \gls{knn}. As an example, in the $1.5\times$ scenario, \seismic with (without) \gls{knn} is $3.0\times$ ($2.3\times$) faster than \kannolo at $97\%$ accuracy cutoff.  The speedups increase to $6.9\times$ ($4.4\times$) in the $2\times$ scenario at $98\%$ accuracy cutoff.

Observe that the \gls{knn} has a notable impact on memory: Storing it costs $ (\lfloor \log_{2} (N-1)\rfloor  + 1)N\kappa$ bits, with $N$ denoting the size of the collection, which translates to $27$ bits per document in our setup. This analysis suggests that reducing the memory impact of the \gls{knn}, for example, using delta encoding to store the ids of the documents, may help \seismic become even faster at a given memory budget.

\subsubsection{Index construction time and size}
\seismic builds the \msmarco index much more quickly than other methods as previously shown in Table~\ref{table:results-rq2}. This advantage holds strong on \msmarcodos, as we show in Table~\ref{table:indexing-time}: \seismic (without the \gls{knn}) builds its index in about $30$ minutes, while \kannolo in the sparse domain takes about $12.5$ hours. The reported configuration refers to the $1.5\times$ memory budget scenario.

Enabling the \gls{knn} in \seismic substantially increases the indexing time. That is to be expected. Despite the construction of the \gls{knn} being approximated with \seismic itself, this process involves the daunting task of searching through approximately $140$ million points. Moreover, as the collection size grows, not only would one have more queries to search, but there would be a larger number of documents to search over. As such, the inference benefits of the \gls{knn} shown in Figure~\ref{fig:k10} come at a high cost at indexing time.

\begin{table}[b]
\caption{Index size and build time for winning configurations.}
 \label{table:indexing-time}
	\centering
\adjustbox{max width=\textwidth}{
	\begin{tabular}{l|@{\hspace{15pt}}l@{\hspace{15pt}}r @{\hspace{15pt}}r}
    \toprule
    Model & Configuration & Index size (GB)   & Build time (hours) \\
    \midrule
    \kannolo & M=32 &104.4  &  10.1 \\
    \seismic (with \gls{knn}) & $\lambda=4e4$, $\kappa =20$& 98.4 & 16.6 \\
    \seismic (without \gls{knn}) &$\lambda=6e4$ & 98.5 & 0.5\\
    
     \bottomrule
	 \end{tabular}
}
\end{table}

\subsection{Scaling Laws}
\begin{figure}[!t]
    \centering
    \includegraphics[width=0.85\linewidth]{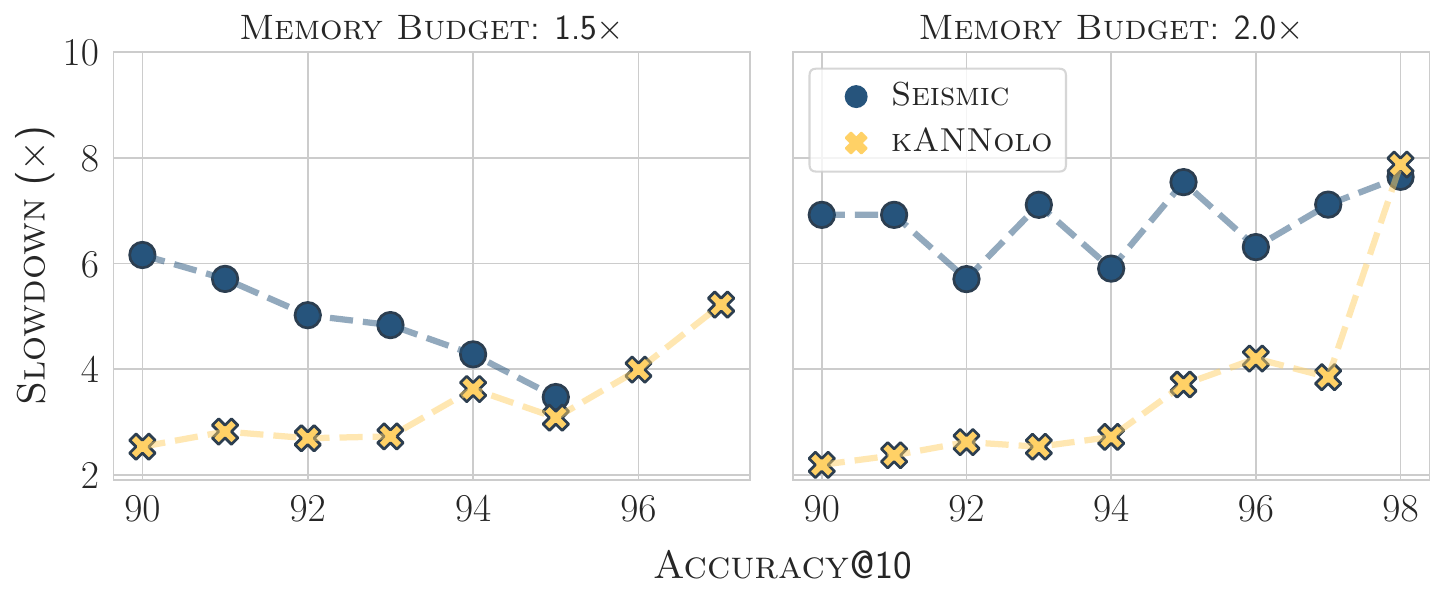}
    \caption{Scaling laws of \seismic and sparse \kannolo. For each accuracy cutoff, we measure the ratio between the latency of a method on \msmarcodos and on \msmarco.
    }
    \label{fig:scaling}
\end{figure}

In Figure~\ref{fig:scaling}, we present the search time scaling laws for \seismic and \kannolo. For each accuracy threshold on the horizontal axis, we plot on the vertical axis the ratio of the search time on \msmarcodos to search time on \msmarco. In doing so, we report the amount of slowdown between the two datasets.

Considering the fact that \msmarcodos is approximately $15\times$ larger than \msmarco, from Figure~\ref{fig:scaling} we learn that both methods scale effectively with dataset size. Concerning \seismic, all configurations show a slowdown of less than $8\times$, except for accuracy thresholds of $97$ and $98$. Regarding \kannolo, in the $1.5\times$ memory budget scenario, slowdowns are reduced, ranging from $2.5\times$ to $5.2\times$. In the $2\times$ memory budget, the slowdown is small for mid accuracy cuts ($90$-$94$) and then, as with \seismic, it increases, reaching a peak of $7.9\times$ at $98\%$ accuracy.

\section{Concluding Remarks} \label{sec:conclusions}

We presented \seismic, a novel approximate algorithm that facilitates effective and efficient retrieval over sparse embedding vectors. We showed empirically its remarkable efficiency on a number of embeddings of publicly-available datasets. \seismic outperforms existing methods, including the winning algorithms at the BigANN Challenge in NeurIPS 2023 that use similar-sized (or larger) indexes. Incredibly, one configuration of \seismic performs top-$10$ \gls{anns} over nearly $140$ million vectors with almost exact accuracy within just $3$ milliseconds.

The efficiency of \seismic is in large part due to a sketching algorithm we designed specifically for sparse vectors. As we explained the evolution of the sketch, dubbed \gls{setsketch}, and presented theoretical results, the sketch can dramatically reduce the number of nonzero coordinates of a nonnegative sparse vector while preserving relative pairwise distances with arbitrarily high probability. That enabled us to reduce the size our novel inverted index, which in turn leads to fewer data points being evaluated.

\vspace{1mm} \noindent \textbf{Acknowledgements}. This work was partially supported by the Horizon Europe RIA ``Extreme Food Risk Analytics'' (EFRA), grant agreement n. 101093026, by the PNRR - M4C2 - Investimento 1.3, Partenariato Esteso PE00000013 - ``FAIR - Future Artificial Intelligence Research'' - Spoke 1 ``Human-centered AI'' funded by the European Commission under the NextGeneration EU program, by the PNRR ECS00000017 Tuscany Health Ecosystem Spoke 6 ``Precision medicine \& personalized healthcare'' funded by the European Commission under the NextGeneration EU programme, by the MUR-PRIN 2017 ``Algorithms, Data Structures and Combinatorics for Machine Learning'', grant agreement n. 2017K7XPAN\_003, and by the MUR-PRIN 2022 ``Algorithmic Problems and Machine Learning'', grant agreement n. 20229BCXNW.

\bibliographystyle{ACM-Reference-Format}
\bibliography{biblio}

\end{document}